\documentstyle[amssymb,aps,twoside]{revtex}  

\begin{document}

\title{(4+$N$)-Dimensional Elastic Manifolds in Random Media:\\ A
Renormalization-Group Analysis}

\author{H.~Bucheli$^a$, O.S.~Wagner$^a$, V.B.~Geshkenbein$^{a,\, b}$,\\
A.I.~Larkin$^{b,c}$,\ and G.~Blatter$^a$\ }

\address{$^{a\,}$Theoretische Physik, ETH-H\"onggerberg, CH-8093 Z\"urich,
Switzerland} 
\address{$^{b\,}$L. D. Landau Institute for Theoretical Physics, 117940
Moscow, Russia} 
\address{$^{c\,}$Theoretical Physics Institute, University of Minnesota,
Minneapolis, MN 55455}

\date{August 1997}

\maketitle
\vspace{1truecm}

\begin{abstract}

Motivated by the problem of weak collective pinning of vortex lattices in
high-temperature superconductors, we study the model system of a
four-dimensional elastic manifold with $N$ transverse degrees of freedom
(4+$N$-model) in a quenched disorder environment. 
We assume the disorder to be weak and short-range correlated, and neglect
thermal effects. 
Using a real-space functional renormalization group (FRG) approach, we derive 
a RG equation for the pinning-energy correlator up to two-loop correction. 
The solution of this equation allows us to calculate the size $R_c$ of
collectively pinned elastic domains as well as the critical force $F_c$, i.e.,
the smallest external force needed to drive these domains. 
We find $R_c\propto\delta_p^{\alpha_2}\,\exp(\alpha_1/\delta_p)$ and
$F_c\propto\delta_p^{-2\alpha_2}\,\exp(-2\alpha_1/\delta_p)$, where $\delta_p\ll 1$
parametrizes the disorder strength, $\alpha_1=(2/\pi)^{N/2}\,8\pi^2/(N+8)$, and
$\alpha_2=2(5N+22)/(N+8)^2$. 
In contrast to lowest-order perturbation calculations which we briefly
review, we thus arrive at determining both $\alpha_1$ (one-loop) and $\alpha_2$
(two-loop).

\end{abstract}


\section{Introduction}

The generic problem of a $D$-dimensional elastic manifold with $N$ transverse
degrees of freedom subject to quenched random impurities that can pin the
manifold has attracted a great deal of attention for many years. It has
applications in numerous areas of physics such as dislocations, spin systems,
polymers, charge density waves, and vortex lattices in type-II
superconductors. The case $N=1$, for instance, corresponds to $D$-dimensional
interfaces separating two coexisting phases in $(D+1)$-dimensional systems. On
intermediate distances the flux lattice in superconductors is believed to
behave like an elastic manifold with $D=3$ and $N=2$.
 
The crucial quantity for defining order in these systems is the
disorder-averaged square of the relative displacement $\langle
u^2({\bf x})\rangle:=\langle [{\bf u}({\bf x})-{\bf u}({\bf 0})]^2\rangle$,
which describes the roughening of the $N$-component displacement field ${\bf
u}({\bf x})$ of $D$-dimensional support. In general, $\langle u^2({\bf
x})\rangle$ depends on the distance $R:=|{\bf x}|$, the dimensionalities $D$
and $N$, and on the type of disorder. In the present paper, we choose the
disorder to be weak, short-range correlated, and of Gaussian type. Under these
assumptions it has been shown\cite{l} that in less than four dimensions ($D\le
4$) quenched disorder always destroys the translational long-range order of
the manifold, breaking it up into correlated domains of extent $R_c$. Each of
these domains behaves elastically independently and is pinned
individually. The length scale $R_c$ over which short-range order subsists is
commonly referred to as the {\em collective pinning radius} and is usually
defined by the criterion $\langle u^2(R_c)\rangle\sim\xi^2$, where $\xi$ is
the length scale describing the internal structure of the
manifold\cite{l,bfglv}. For distances $R$ smaller than $R_c$ (perturbative
regime), the ground state of the system is unique and the mean square
displacement $\langle u^2 (R)\rangle$ can be calculated perturbatively. On the
other hand, for $R>R_c$ (random manifold regime), there are many competing
metastable minima inducing the so-called ``wandering'' of the manifold as
described by the scaling law  $\langle u^2
(R)\rangle\approx\xi^2\,(R/R_c)^{2\zeta}\,(R\gg R_c)$.   
A lot of effort has gone into the determination of the wandering exponent
$\zeta$, using elaborate techniques such as the replica formalism combined
with either variational approaches\cite{mp1,mp2} or with the renormalization
group\cite{f,bf,bbm,h1,h2} (RG). Here, we make use of RG methods within the
perturbative regime to determine the collective pinning radius $R_c$. For that
purpose, we will replace the estimate $\langle u^2(R_c)\rangle\sim\xi^2$ by an
improved definition of $R_c$ based on the divergence of the fourth derivative
of the pinning-energy correlator, signaling the appearance of competing
ground states at this length scale.     
The collective pinning radius $R_c$ is not only a relevant quantity for
characterizing order, but also determines dynamic properties such as the
minimum force $F_c$ needed to move the pinned manifold (critical force) and
the activation barrier $U_c$ for creep. 
 
In this work we concentrate on the particularly interesting case $D=4$, the
upper critical dimension of this problem, where the mean-square
displacement shows a logarithmic behavior in the perturbative regime\cite{rem1},
$\langle u^2(R)\rangle\propto\ln R,\,(R<R_c)$. This logarithmic situation
motivates the use of renormalization group methods. The (4+$N$)-dimensional
model system is strongly related to the problem of weak collective pinning of a
(3+2)-dimensional manifold with dispersive elastic moduli, an issue which has
received a lot of interest in connection with pinning and creep behavior of
vortices in high-$T_c$ superconductors. In a future paper\cite{wglb}, the RG
technique developed below will be applied to that topic (see also Ref.~[19]).    

The main objects of interest in our investigations are the collective pinning
radius $R_c$ and the critical force density $F_c$. 
In section II, we introduce the non-dispersive (4+$N$)-model and define the
static and dynamic Green's functions as well as the relevant correlation
functions. For simplicity, we neglect thermal effect, i.e., we set $T=0$. Then,
in section III, we derive $R_c$ and $F_c$ by means of lowest-order
perturbation calculations combined with scaling techniques and the dynamic approach,
and discuss the problem with these methods. Section IV is devoted to the
renormalization group treatment of the pinning problem. We construct the RG
transformation and derive a functional RG equation for the pinning energy
correlator in one and two-loop approximation. We show how $R_c$ can be
obtained by solving this equation and determine the critical force density
$F_c$ using simple scaling relations. We discuss various aspects making this
technique superior to those presented in section III. In a second step, we apply
these results to the dynamical situation, where the system is driven with an
external force. In particular, we determine the behavior of the friction
coefficient $\eta$ under the action of the RG and give an alternative
derivation of $F_c$. Finally, in section VI we summarize and present our conclusion.

\section{Model}

\subsection{Non-dispersive (4+$N$)-model}

We consider a four-dimensional elastic manifold in a
(4+$N$)-dimensional space in the continuum limit: A point of the manifold is
represented by a four-component vector ${\bf x}$ ({\em internal} degrees of
freedom), while the displacement relative to the equilibrium position at that
point is characterized by the continuous $N$-component vector field
${\bf u}({\bf x})$ ({\em transverse} degrees of freedom). The elastic part of the
free energy is given by the isotropic expression
$${\cal F}_{el}[{\bf u}]=\frac{1}{2}\,C\int d^4 x\,\nabla u^\alpha({\bf
x})\cdot\nabla u^\alpha({\bf x})\,,$$ 
where $\alpha=1,2,\dots,N$; as usual, indices appearing twice are implicitly
summed over. In the present work, the elastic modulus $C$ is taken to be a
constant (non-dispersive model).

The elastic manifold is embedded in a random environment which we model as a
Gaussian random pinning potential $U_{pin}({\bf x},{\bf s})$ with zero mean and
short-range correlations, 
\begin{equation} \langle U_{pin}({\bf x},{\bf s})\,U_{pin}({\bf x}',{\bf
s}')\rangle = \gamma_U\,\delta^4({\bf x}-{\bf x}')\,\delta^N ({\bf s}-{\bf
s}')\,.\label{UUcorr}\end{equation} 
The parameter $\gamma_U$ is a measure of the disorder strength and is assumed
to be small. Angular brackets $\langle\dots\rangle$ denote the average
over all possible realizations of disorder. Note that $U_{pin}$ depends on the
internal as well as on the transverse degrees of freedom of the system.
For a given configuration ${\bf u}({\bf x})$ and a random, but fixed disorder, the
energy density arising from the interaction of the manifold with the
disordered potential is the convolution of the potential $U_{pin}$ with the
form factor $p(s)$:
\begin{equation} E_{pin}({\bf x},{\bf u}({\bf x}))=\int
d^N\! s\,U_{pin}({\bf x},{\bf s})\,p(|{\bf s}-{\bf u}({\bf x})|)\,.
\label{Epin}\end{equation} 
The function $p(s)$ characterizes the internal structure of the manifold. In
the following, we will adopt the simple expression  
\begin{equation} p(s)=\exp\left(-\frac{s^2}{\xi^2}\right)\,,\label{formfactor}
\end{equation}
where $\xi$ is chosen as the smallest resolvable length scale of the
system. The effect of the form factor is to smear the random potential
$U_{pin}$ over a length $\xi$, so that the typical effective distance between
two `valleys' in the energy landscape is of order $\xi$.

The total free energy of the manifold in the presence of disorder is
eventually  
\begin{equation}{\cal F}[{\bf u}]={\cal F}_{el}[{\bf u}]+{\cal F}_{pin}[{\bf
u}]=\int d^4 x\,\left[\frac{1}{2}C\nabla u^\alpha ({\bf x})\cdot\nabla u^\alpha
({\bf x})+E_{pin}({\bf x},{\bf u}({\bf x}))\right]\,,\label{Ftot}\end{equation}
with the pinning energy correlator
\begin{equation}\langle E_{pin}({\bf x},{\bf u}({\bf x}))\,E_{pin}({\bf x}',{\bf
u}'({\bf x}'))\rangle =\delta^4({\bf x}-{\bf x}')\,K_\xi ({\bf u}({\bf
x})-{\bf u}'({\bf x}'))\,,\label{EEcorr}\end{equation} 
where
\begin{equation}K_\xi ({\bf
u})=\gamma_U\,\xi^N\,\left(\frac{\pi}{2}\right)^{\frac{N}{2}}\,\exp\left(-\frac{u^2}{2\xi^2}\right)\,,\quad
u:=|{\bf u}|\,,\label{K}\end{equation}
as obtained from Eqs.~(\ref{UUcorr})--(\ref{formfactor}). The exponential in
the correlator $K_\xi ({\bf u})$ is directly related to the expression for the
form factor, Eq.~(\ref{formfactor}).     

The equilibrium configuration ${\bf u}({\bf x})$ of the system results from
the competition between the elastic energy the system has to pay for a
distortion and the pinning energy the system can win by accommodating to the
impurity potential. Unfortunately, determining these configurations reveals an
impossible undertaking because of the randomness of $E_{pin}$. One must
therefore content oneself with calculating disorder-averaged quantities,
relying on the system's assumed property of self-averaging.

\subsection{Green's functions}

Minimizing the free-energy functional (\ref{Ftot}) with respect to ${\bf
u}({\bf x})$, 
\begin{equation}\frac{\delta}{\delta u^\alpha ({\bf x})}{\cal F}[{\bf
u}]=0\,,\quad\alpha=1,\dots,N\,,\label{Fmin}\end{equation}
defines the equation of state,
\begin{equation}-C\,\nabla^2 u^\alpha ({\bf x})=F_{pin}^\alpha ({\bf x},{\bf
u}({\bf x}))\,,\label{eqstate}\end{equation}
where ${\bf F}_{pin}$ stands for the pinning force density,
\begin{equation}
F_{pin}^\alpha ({\bf x},{\bf u}({\bf x})):=-\frac{\delta}{\delta u^\alpha ({\bf
x})}{\cal F}_{pin}[{\bf u}]=-\frac{\partial}{\partial u^\alpha} E_{pin}({\bf x},{\bf
u}({\bf x}))\,.\label{Fpin}\end{equation} 
%
%Note that with this procedure we implicitly assume that the ground state of
%the system is unique.
The equation of state (\ref{eqstate}) can equivalently be written in the form
\begin{equation}u^\alpha({\bf
x})=\int\,d^4y\,G^{\alpha\beta}({\bf x}-{\bf y})\,F_{pin}^\beta ({\bf y},{\bf
u}({\bf y}))\,,\label{inteq}\end{equation} 
where
\begin{equation} G^{\alpha\beta}({\bf x})=\delta^{\alpha\beta}\,G({\bf
x})\,,\quad G({\bf x})=\frac{1}{(2\pi)^2 Cx^2}\,,\quad
x:=|{\bf x}|\,,\label{Greenstat}\end{equation}    
is the static Green's function of the system defined by the relation 
$$-C\,\nabla^2 G({\bf x})=\delta^4({\bf x})\,.$$
 
For the dynamic approach in section III we need the time-dependent Green's
function $G^{\alpha\beta}({\bf x},t)$. We assume a dissipative dynamics for our
elastic manifold  which we characterize by the friction coefficient
$\eta_\circ$. With the equation of motion 
$$\left(-C\,\nabla^2+\eta_\circ\,\frac{\partial}{\partial t}\right)
u^\alpha ({\bf x},t)=F_{pin}^\alpha ({\bf x},{\bf u}({\bf x},t))\,,$$  
the dynamic Green's function takes the form $\hat{G}^{\alpha\beta}({\bf
k},\omega)=\delta^{\alpha\beta}/(Ck^2-i\eta_\circ\omega)$ in Fourier representation.
Transforming back to real space, one readily obtains
\begin{equation} G^{\alpha\beta}({\bf x},t)=\delta^{\alpha\beta}\,G({\bf
x},t)\,,\quad G({\bf x},t)=\frac{\Theta(t)\,\eta_\circ}{16\pi^2 C^2
t^2}\,\exp\left(-\frac{\eta_\circ x^2}{4Ct}\right)\,.\label{Greendyn}\end{equation}
The Heaviside-function $\Theta(t)$ reflects the causality of the Green's function.

\subsection{Correlation functions}

For later use we define the derivatives of $E_{pin}$, 
$$E_{pin}^{\alpha_1\cdots\alpha_l}({\bf x},{\bf u}({\bf
x})):=\frac{\partial}{\partial u^{\alpha_1}}\cdots\frac{\partial}{\partial
u^{\alpha_l}}E_{pin}({\bf x},{\bf u}({\bf
x}))\,,\quad\alpha_1,\dots,\alpha_l=1,\dots N\,,\quad l=1,2,3,\dots\,,$$ 
and the corresponding correlators,
\begin{equation} \langle
E_{pin}^{\alpha_1\cdots\alpha_l}({\bf x},{\bf u}({\bf
x}))\,E_{pin}^{\beta_1\cdots\beta_n}({\bf x}',{\bf u}'({\bf x}'))\rangle
=(-1)^n\,\delta^4({\bf x}-{\bf
x}')\,K_\xi^{\alpha_1\cdots\alpha_l\beta_1\cdots\beta_n}({\bf u}({\bf x})-{\bf
u}'({\bf x}'))\,,\label{dercorr}\end{equation} 
where naturally $K_\xi^{\alpha\beta\cdots}({\bf u}):=(\partial/\partial
u^{\alpha})(\partial/\partial u^{\beta})\cdots K_\xi ({\bf u})$. Each partial
derivative $\partial/\partial u'^{\,\alpha}$ contributes a minus sign, thus
giving rise to the factor $(-1)^n$. In particular, the force-force correlator
is given by 
\begin{equation}K_\xi^{\alpha\beta}({\bf
u})=-\gamma_U\,\xi^{N-2}\,\left(\frac{\pi}{2}\right)^{\frac{N}{2}}\,\left(\delta^{\alpha\beta}-\frac{u^\alpha
u^\beta}{\xi^2}\right)\,\exp\left(-\frac{u^2}{2\xi^2}\right)\,.\label{K2}\end{equation}
Of special interest in section IV will be the curvature of the force at the origin,
\begin{equation}K_\xi^{\alpha\beta\gamma\delta}(0)=\Gamma_\xi\,\left(\delta^{\alpha\beta}\delta^{\gamma\delta}+\delta^{\alpha\gamma}\delta^{\beta\delta}+\delta^{\alpha\delta}\delta^{\beta\gamma}\right)=:\Gamma_\xi\,\Delta^{\alpha\beta\gamma\delta}\,,\label{K4}\end{equation}
with $\Gamma_\xi:=\gamma_U\,\xi^{N-4}\,\left(\frac{\pi}{2}\right)^{\frac{N}{2}}$.

\section{Perturbation theory}

In this section we present the usual ways to define and determine the
collective pinning radius $R_c$ and the critical force density $F_c$. 
First, we calculate $R_c$ by a lowest-order perturbation calculation, and
assess $F_c$ by dimensional estimates \cite{lo2}. Second, we use the
dynamic approach \cite{hs,lo1} to determine $F_c$ in an alternative manner. We
also point out the disadvantages of these simple methods.

\subsection{Dimensional estimates}

The collective pinning radius $R_c$ is usually derived by computing the
fluctuations of the displacement field ${\bf u}({\bf x})$ induced by the
random environment, 
$$\langle u^2(R)\rangle:=\langle [{\bf u}({\bf x})-{\bf u}({\bf
0})]^2\rangle\,,\quad R:=|{\bf x}|\,,$$ 
in the absence of an external force, ${\bf F}_{ext}=0$. The condition $\langle
u^2(R_c)\rangle\simeq\xi^2$ defines the collective pinning radius $R_c$ which
represents the extension of a collectively pinned elastic domain. For small
fluctuations, $\langle u^2(R)\rangle\le\xi^2$, one can use a perturbative
approach\cite{lo2}. The starting point is the integral equation
(\ref{inteq}). We expand the pinning force, ${\bf F}_{pin}({\bf y},{\bf
u}({\bf y}))={\bf F}_{pin}({\bf y},0)+O({\bf u}({\bf y}))$, and only retain
the lowest-order term, so that Eq.~(\ref{inteq}) reads
($G^{\alpha\beta}=\delta^{\alpha\beta}G$)  
\begin{equation}u^\alpha ({\bf x})=\int\,d^4y\,G({\bf x}-{\bf
y})\,F_{pin}^\alpha ({\bf y},0)\,.\label{inteqlop}\end{equation} 
From this expression and by means of Eqs.~(\ref{Fpin}) and (\ref{dercorr}), we
obtain
$$\langle u^\alpha ({\bf x}) u^\alpha ({\bf x}')\rangle =
-K_\xi^{\alpha\alpha}(0)\,\int d^4 y\,G({\bf x}-{\bf y})\,G({\bf
x}'-{\bf y})+O(\gamma_U^2)\,.$$    
Next, we write the Green's functions in Fourier representation; the mean
square of the relative displacement then takes the form
\begin{equation}\langle u^2(R)\rangle = -2 K_\xi^{\alpha\alpha}(0)\,\int\frac{d^4
k}{(2\pi)^4}\,\Big(1-\cos({\bf k}\cdot{\bf x})\Big)\,\hat{G}({\bf
k})^2+O(\gamma_U^2)\,.\label{msdlop}\end{equation} 
Surprisingly enough, this lowest-order perturbation expression is true to all
orders in $\gamma_U$ \cite{el} (this can be checked by keeping ${\bf u}({\bf y})$ on
the rhs.\ of Eq.~(\ref{inteqlop}) and going through the subsequent
calculations). Two elements are responsible for this striking fact: the
short-range correlation $\propto\delta^4 ({\bf x}-{\bf y})$ in $\langle
E_{pin}\,E_{pin}\rangle$ and the implicit assumption made in this derivation
that the system is in a unique state, which is only valid in the perturbative
regime. The integral in Eq.~(\ref{msdlop}) is invariant under space rotations;
one can therefore choose ${\bf x}$ parallel to the $x_4$-axis and introduce
cylindrical coordinates ($K^2:=k_1^2+k_2^2+k_3^2$),
$$\langle u^2(R)\rangle=-\frac{2 K_\xi^{\alpha\alpha}(0)}{(2\pi)^4
C^2}\,\int_0^\infty 4\pi K^2 dK\int_{-\infty}^\infty dk_4\,\frac{1-\cos(k_4
R)}{(K^2+k_4^2)^2}\,.$$
Performing the $K$-integration yields
\begin{equation}\langle u^2(R)\rangle=-\frac{K_\xi^{\alpha\alpha}(0)}{4\pi^2
C^2}\int_0^\infty dk_4\,\frac{1-\cos(k_4 R)}{k_4}\,.\label{intk4}\end{equation}
The $k_4$-integration produces a log-contribution which diverges for
$k_4\rightarrow\infty$. For this reason, one has to introduce an upper cutoff
given by the inverse of the smallest length scale in the problem which is
$\xi$. With $K_\xi^{\alpha\alpha}(0)=-\gamma_U\,\xi^{N-2}\,N (\pi/2)^{N/2}$,
Eq.~(\ref{K2}), one finally obtains    
\begin{equation}\langle
u^2(R)\rangle\approx
\frac{N}{4\pi^2}\,\left(\frac{\pi}{2}\right)^{\frac{N}{2}}\delta_p\,\ln\left(\frac{R}{\xi}\right)\,,\quad
R\gg\xi\,,\label{u2alo}\end{equation}
where $\delta_p:=\gamma_U\,\xi^{N-4}/C^2$ is the dimensionless disorder
parameter. The collective pinning radius defined through the relation $\langle
u^2(R_c)\rangle\simeq\xi^2$ then is 
\begin{equation}R_c\simeq\xi\,\exp\left(\frac{4\pi^2}{N}\left(\frac{2}{\pi}\right)^{\frac{N}{2}}\frac{1}{\delta_p}\right)\,,\label{Rclop}\end{equation}
which is exponentially large in the limit of weak pinning, $\delta_p\ll 1$.
With the upper cutoff in Eq.~(\ref{intk4}) being given only up to a factor of order
unity, the constant of proportionality in Eq.~(\ref{Rclop}) is not
unequivocally determined. Much more relevant, though, is the uncertainty in
the numerical factor in the exponential function, which has its origin in the
criterion $\langle u^2(R_c)\rangle\simeq\xi^2$, where the constant of
proportionality is assumed to be of order unity. 

A rough estimate for the critical force density $F_c$ can be gained by
scaling. The typical elastic energy of a collectively pinned domain is 
$$U_c\sim C\,\left(\frac{\xi}{R_c}\right)^2\,V_c\,,$$
where $V_c:=R_c^4$ is the collective pinning volume. The energy gain arising
from the action of an external force $F_{ext}$ shifting the domain by $\xi$ is
$$U_{ext}\sim F_{ext}\,V_c\,\xi\,.$$
Comparison between both energy scales gives a scaling estimate for the
critical force density, 
\begin{equation}
F_c\sim\frac{C}{\xi}\left(\frac{\xi}{R_c}\right)^2\;\simeq\frac{C}{\xi}\,\exp\left(
-\frac{8\pi^2}{N}\left(\frac{2}{\pi}\right)^{\frac{N}{2}}\frac{1}{\delta_p}\right)\,.\label{Fcdimest}\end{equation}

\subsection{Dynamic approach}

In the dynamic approach\cite{hs,lo1} the critical force $F_c$ is determined
directly without any reference to the pinning radius $R_c$. To begin 
with, the system is driven by a large external force density $F_{ext}\gg
F_c$. In this regime, the elastic manifold does not noticeably feel the
pinning potential and ${\bf F}_{ext}$ is only opposed by the friction force ${\bf
F}_{frict}$ which we describe by the phenomenological expression
${\bf F}_{frict}=-\eta_\circ\,{\bf v}$, with the viscosity $\eta_\circ$ a
material constant. The flow velocity ${\bf v}$ of the manifold is determined by the
steady-flow condition
$${\bf F}_{frict}+{\bf F}_{ext}=0\qquad\Longrightarrow\qquad
{\bf v}={\bf v}_\circ:=\frac{1}{\eta_\circ}\,{\bf F}_{ext}\,.$$
If ${\bf F}_{ext}$ is decreased, the system begins to sensibly interact with
the randomly distributed impurities via the pinning force ${\bf
F}_{pin}$. This becomes noticeable through fluctuations in the velocity field,
leading to a reduction of the average velocity, ${\bf v}={\bf v}_\circ-\delta
{\bf v}$ with ${\bf v}_\circ\cdot\delta {\bf v}>0$. The average of ${\bf F}_{pin}$
can be interpreted as an effective friction force with a velocity-dependent
viscosity $\delta\eta (v)$, 
\begin{equation}\langle {\bf F}_{pin}\rangle=-\delta\eta(v)\,{\bf
v}\,.\label{avFpin}\end{equation} 
As before, the steady-flow velocity is given by the relation
\begin{equation}{\bf F}_{frict}+\langle {\bf F}_{pin}\rangle+{\bf
F}_{ext}=0\quad\Longrightarrow\quad\eta(v)\,{\bf v}:=\big(\eta_\circ+\delta\eta(v)
\big)\,{\bf v}={\bf F}_{ext}\,.\label{flow}\end{equation}
The applied force for which the relative fluctuations of the viscosity are of
order unity then furnishes a useful criterion for the critical force density:
\begin{equation}\left.\frac{\delta\eta(F_{ext})}{\eta_\circ}\right|_{F_{ext}=F_c}=1\,.\label{critFc}\end{equation}
Alternatively, a criterion can be formulated making use of the
reduction in the flow velocity, $\langle {\bf F}_{pin}\rangle=-\eta_\circ
\delta {\bf v}$, and the condition $\delta v(v_c)=v_c$.   

We now calculate the friction coefficient $\delta\eta$ due to pinning within a
perturbative approach up to first order in $\delta_p$. We start from
Eq.~(\ref{avFpin}),
$$v^\alpha\,\delta\eta=-\langle\,F_{pin}^\alpha({\bf x},{\bf v}t+{\bf u}({\bf
x},t))\,\rangle\,,$$ 
and expand the pinning force into a Taylor series, keeping the first two
terms,
$$v^\alpha\,\delta\eta\simeq -\langle F_{pin}^\alpha ({\bf x},{\bf
v}t)\rangle-\langle F^{\alpha\beta}_{pin}({\bf x},{\bf v}t)\,u^\beta ({\bf
x},t)\rangle\,.$$ 
The first term vanishes; in the second one, we use the integral equation
(\ref{inteq}) generalized to the time-dependent case and expanded to lowest order,
$$u^\alpha({\bf x},t)=\int d^4y\,ds\,G({\bf x}-{\bf
y},t-s)\,F_{pin}^\alpha({\bf y},{\bf v}s)\,.$$ 
Remembering Eq.~(\ref{dercorr}) and the definition
$F_{pin}^\alpha=-E_{pin}^\alpha$, one obtains 
\begin{equation}v^\alpha\delta\eta\simeq\int dt \,G({\bf
0},t)\,K_\xi^{\alpha\beta\beta}({\bf v}t)\,,\label{dyn1}\end{equation}
where we made use of the time-translation symmetry. Inserting the explicit
expression for the correlator $K_\xi^{\alpha\beta\beta}$ and for the time-dependent
Green's function (\ref{Greendyn}), we find:
$$\frac{\delta\eta(v)}{\eta_\circ}\simeq\frac{N+2}{16\pi^2}\left(\frac{\pi}{2}\right)^{\frac{N}{2}}\,\delta_p\int_0^\infty
\frac{dt}{t}\,\exp\left(-\frac{v^2\,t^2}{2\xi^2}\right)+O(\delta_p^2)\,.$$
The integration produces a log-divergence for $t\rightarrow 0$. A lower cutoff
is given by the time $t_\xi$ the elastic manifold needs to recover from a
distortion on the smallest length scale $\xi$. Balancing elastic and dynamic
terms in the Green's function $G({\bf x},t)$, Eq.~(\ref{Greendyn}), on the
scale $\xi$, we obtain 
\begin{equation}
t_\xi\simeq\frac{\eta_\circ\,\xi^2}{4C}\,.\label{t_xi}\end{equation} 
On the other hand, the exponential factor in the integrand, which originates
from the correlator $K_\xi^{\alpha\beta\beta}$, provides an upper cutoff,
\begin{equation} t_v\simeq\frac{\sqrt{2}\,\xi}{v}\,,\label{t_v}\end{equation}
describing the time scale for the interaction between the system and the
pinning centers at velocity $v$. With these two cutoffs, the ratio
between $\delta\eta$ and $\eta_\circ$ becomes
\begin{equation}\frac{\delta\eta(F_{ext})}{\eta_\circ}\simeq\frac{N+2}{16\pi^2}\,\left(\frac{\pi}{2}\right)^{\frac{N}{2}}\,\delta_p\,\left[\ln\left(\frac{4\sqrt{2}C}{\xi
F_{ext}}\right)+const.\right]+O(\delta_p^2)\,.\label{etalop}\end{equation}
This expression is valid in the non-linear regime, i.e., for forces
$F_{ext}<4\sqrt{2}C/\xi$ (so that $t_\xi<t_v$), where the effective friction
coefficient $\eta(F_{ext})=F_{ext}/v(F_{ext})$ deviates from $\eta_\circ$. The
critical force $F_c$ is found from the result (\ref{etalop}), using the criterion
$\delta\eta(F_c)/\eta_\circ\simeq 1$ (for weak disorder $\delta_p\ll 1$, we
expect $F_c$ to be much smaller than $C/\xi$, so that the constant in
Eq.~(\ref{etalop}) can be neglected). We obtain
\begin{equation}F_c\simeq\frac{4\sqrt{2}C}{\xi}\,\exp\left(-\frac{16\pi^2}{N+2}\,\left(\frac{2}{\pi}\right)^{\frac{N}{2}}\frac{1}{\delta_p}\right)\,,\label{Fcdynap}\end{equation}
which is indeed exponentially small. As in the previous method, both the
constant of proportionality and the numerical factor in the exponential function can
only be ascertained up to a number of order unity.

\section{Renormalization Group Analysis}

As we have seen in the previous section, the leading term in the fluctuations
of the displacement field ${\bf u}({\bf x})$ shows a logarithmic behavior, $\langle
u^2\rangle\propto\ln R^2$, cf.~Eq.~(\ref{u2alo}). The logarithmic dependence
weights all scales equally such that all length scales are relevant in the
final result. This provides the motivation to apply the renormalization group
(RG). Our guiding line for this analysis will be the work of Efetov and
Larkin\cite{el} dealing with the pinning problem of charge-density waves. 

\subsection{Construction of the RG transformation}

The following real-space renormalization procedure was suggested by
Khmel'nitskii and Larkin\cite{kl} for the problem of friction between
two rough surfaces. In a recent publication\cite{bbm}, Balents, Bouchaud, and
M\'{e}zard gave a precise formulation of this method within the framework of
the momentum-shell RG and used it to investigate the random manifold regime
(large scales $R>R_c$) of the ((4-$\epsilon$)+$N$)-model; here, we are
interested in the perturbative regime (small scales $R<R_c$) with the aim to
determine $R_c$.
 
Assume that we have renormalized the displacement up to a scale $R_1>\xi$; let
us denote the corresponding field by ${\bf u}_{(1)}({\bf x})$. The associated
free-energy functional reads
$$ {\cal F}[{\bf u}_{(1)}]=\int d^4 x\,\left[\frac{1}{2}C\nabla u_{(1)}^\alpha
({\bf x})\cdot \nabla u_{(1)}^\alpha ({\bf x})+E_{pin,(1)}({\bf x},{\bf
u}_{(1)}({\bf x}))\right]\,.$$ 
Next, we go over to a larger scale $R_2>R_1$ and separate ${\bf u}_{(1)}({\bf x})$
into a far and a near-field contribution, ${\bf u}_{(1)}({\bf x})={\bf
u}_{(2)}({\bf x})+{\bf w}({\bf x})$, with
\begin{eqnarray} u_{(2)}^\alpha ({\bf x}) & = & -\int\limits_{|{\bf x}-{\bf y}|>R_2}
d^4y\,G^{\alpha\beta}({\bf x}-{\bf y})\,E_{pin,(1)}^\beta ({\bf y},{\bf
u}_{(1)}({\bf y}))\,,\nonumber\\ w^\alpha ({\bf x}) & = & -\int_\Omega
d^4y\,G^{\alpha\beta}({\bf x}-{\bf y})\,E_{pin,(1)}^\beta ({\bf y},{\bf
u}_{(1)}({\bf y}))\,,\label{w}\end{eqnarray}
where $\Omega:=\{{\bf x},{\bf y}|R_1<|{\bf x}-{\bf y}|<R_2\}$. The free energy
can then be written as
$${\cal F}[{\bf u}_{(2)}]=\int d^4 x\,\left[\frac{1}{2}C\nabla u_{(2)}^\alpha
({\bf x})\cdot \nabla u_{(2)}^\alpha ({\bf x})+E_{pin,(2)}({\bf x},{\bf
u}_{(2)}({\bf x}))\right]$$ 
with the renormalized pinning energy density,
$$E_{pin,(2)}({\bf x},{\bf u}_{(2)}({\bf x}))=E_{pin,(1)}({\bf x},{\bf
u}_{(1)}({\bf x}))+\frac{1}{2}C\left[\nabla w^\alpha ({\bf x})\cdot\nabla w^\alpha
({\bf x})+2\,\nabla u_{(2)}^\alpha ({\bf x})\cdot\nabla w^\alpha ({\bf
x})\right]\,.$$ 
The mixed term $\nabla u_{(2)}^\alpha\cdot\nabla w^\alpha$ is zero when
integrated over and can be omitted. Integrating the quadratic term by parts,
one finds 
\begin{equation}E_{pin,(2)}({\bf x},{\bf u}_{(2)}({\bf x}))=E_{pin,(1)}({\bf x},{\bf
u}_{(1)}({\bf x}))-\frac{1}{2}w^\alpha
({\bf x})\,C\nabla^2w^\alpha({\bf x})\,.\label{RGEpin}\end{equation}
In the limit of zero temperature, out of all the possible near-field
contributions ${\bf w}({\bf x})$ only those minimizing the energy are
relevant (in a statistical sense),
$$\frac{\partial E_{pin,(2)}({\bf x},{\bf u}_{(2)}({\bf x}))}{\partial
w^\alpha}=0\,.$$ 
Combining this condition with Eq.~(\ref{RGEpin}), we can relate ${\bf w}({\bf
x})$ to the pinning force $E^\alpha_{pin,(1)}$,
\begin{equation}-C\nabla^2 w^\alpha ({\bf x})=-E_{pin,(1)}^\alpha ({\bf x},{\bf
u}_{(1)}^\alpha ({\bf x}))\,.\label{mincond}\end{equation}
(It is interesting to note that by differentiating Eq.~(\ref{RGEpin}) with
respect to $u_{(2)}^\alpha$, one shows that for the pinning force
$$E_{pin,(2)}^\alpha ({\bf x},{\bf u}_{(2)}({\bf x}))=E_{pin,(1)}^\alpha ({\bf
x},{\bf u}_{(1)}({\bf x}))+O\left(\frac{\partial w}{\partial u_{(2)}}\right)\,,$$
holds, which tells us that the force correlator $K^{\alpha\beta}({\bf u}=0)$
will not change under the RG.)

Eq.~(\ref{RGEpin}), together with Eqs.~(\ref{w}) and (\ref{mincond}), is the
starting point for our further considerations. It will allow us to derive a
functional renormalization group (FRG) equation for the pinning energy
correlator $K_R ({\bf u})$, the subscript $R$ denoting the scale of
renormalization. We will then solve this equation by expanding the relevant
correlators around ${\bf u}=0$. It turns out that the fourth derivative,
$K^{\alpha\beta\gamma\delta}_R(0)$, as well as higher-order even derivatives diverge
at a {\it finite} scale which we identify with the collective pinning radius
$R_c$. Indeed, we know that on length scales smaller than $R_c$ (perturbative
regime), the system is in a unique state and the pinning energy as well as the
related correlators are analytic functions in the field ${\bf u}$. At $R_c$
the manifold starts to probe different energy `valleys' and our calculation,
which rests on the analyticity of the pinning energy, breaks down, reflected
by the singular behavior of the curvature of $K({\bf u})$. In what follows we
will take the divergence of $K_R^{\alpha\beta\gamma\delta}(0)$ as an
unambiguous definition of $R_c$ which allows one to go beyond the simple
estimate $\langle u^2(R_c)\rangle\simeq\xi^2$.   

\subsection{One-loop FRG equation}

We first construct a formal expansion in ${\bf w}$ of $E_{pin,(2)}$,
Eq.~(\ref{RGEpin}), which we will then use to derive a perturbation series in
$\delta_p$ of $\langle E_{pin,(2)} E_{pin,(2)}\rangle$. Thereby, we will
relate the energy correlators at renormalization scales $R_2$ and $R_1$,
$K_{R_2}$ and $K_{R_1}$, respectively; this will directly lead to the FRG
equation for $K_R ({\bf u})$. To simplify the notation, we will omit the subscript
`pin' and use the abbreviations 
$$\begin{array}{l@{ := }l@{\hspace{1cm}}l@{ := }l}
\bar{E}_j & E_{pin,(2)}({\bf x}_j,{\bf u}_{(2)}({\bf x}_j))\,, &
E_j^{\alpha\beta\cdots} & E_{pin,(1)}^{\alpha\beta\cdots}({\bf x}_j,{\bf
u}_{(2)}({\bf x}_j))\,,\\ 
\bar{E}'_j & E_{pin,(2)}({\bf x}_j,{\bf u}'_{(2)}({\bf x}_j))\,, &
E'^{\alpha\beta\cdots}_j & E_{pin,(1)}^{\alpha\beta\cdots}({\bf x}_j,{\bf
u}'_{(2)}({\bf x}_j))\,,\\
\bar{K}_{ij} & K_{R_2}({\bf u}_{(2)}({\bf x}_i)-{\bf u}'_{(2)}({\bf x}_j))\,,
& K_{ij}^{\alpha\beta\cdots} & 
K_{R_1}^{\alpha\beta\cdots}({\bf u}_2({\bf x}_i)-{\bf u}'_2({\bf
x}_j))\,.\end{array}$$ 
Let us furthermore define the operator
$$G_{ij}^{\alpha\beta}(\cdots):=\int_\Omega
d^4x_j\,G^{\alpha\beta}({\bf x}_i-{\bf x}_j)\,(\cdots)\,.$$ 
We begin by expanding the first term on the rhs.\ of Eq.~(\ref{RGEpin}) with
respect to ${\bf w}({\bf x})$ into a power series:
$$E_{(1)}({\bf x}_0,{\bf u}_{(1)}({\bf x}_0))=E_{(1)}({\bf x}_0,{\bf u}_{(2)}({\bf x}_0)+{\bf w}({\bf x}_0))=\sum_{l=0}^{\infty}\,\frac{1}{l!}\,E_0^{\alpha_1\cdots\alpha_l}\,w^{\alpha_1}({\bf x}_0)\cdots
w^{\alpha_l}({\bf x}_0)\,.$$
Next, we replace everywhere ${\bf w}$ by its implicit definition, 
Eq.~(\ref{w}). The force function $E_{(1)}^\beta$ under the integral is in
turn written as a Taylor expansion as above. We proceed iteratively, thereby
eliminating ${\bf w}$ from our equation. Grouping the different terms with
increasing number of Green's functions, we arrive at 
\begin{eqnarray*}-E_{(1)}({\bf x}_0,{\bf u}_{(1)}({\bf x}_0)) &=&
-E_0+E^\mu_0\,G_{01}^{\mu\nu}\,E_1^\nu-E_0^\mu\,G_{01}^{\mu\nu}\,E_1^{\nu\rho}\,G_{12}^{\rho\sigma}\,E_2^\sigma-\frac{1}{2!}E_0^{\mu\rho}\,G_{01}^{\mu\nu}\,E_1^\nu\,G_{02}^{\rho\sigma}\,E_2^\sigma+\\
& & +E_0^\mu\,G_{01}^{\mu\nu}\,E_1^{\nu\rho}\,G_{12}^{\rho\sigma}\,E_2^{\sigma\kappa}\,G_{23}^{\kappa\lambda}\,E_3^\lambda+\frac{1}{2!}E_0^\mu\,G_{01}^{\mu\nu}\,E_1^{\nu\rho\kappa}\,G_{12}^{\rho\sigma}\,E_2^{\sigma}\,G_{13}^{\kappa\lambda}\,E_3^{\lambda}+\\
& & +\frac{2}{2!}E_0^{\mu\kappa}\,G_{01}^{\mu\nu}\,E_1^{\nu\rho}\,G_{12}^{\rho\sigma}\,E_2^\sigma\,G_{03}^{\kappa\lambda}\,E_3^{\lambda}+\frac{1}{3!}E_0^{\mu\rho\kappa}\,G_{01}^{\mu\nu}\,E_1^\nu\,G_{02}^{\rho\sigma}\,E_2^\sigma\,G_{03}^{\kappa\lambda}\,E_3^\lambda+\dots\,.\end{eqnarray*}
In a diagrammatic language this can be written as
\vspace{-1cm}
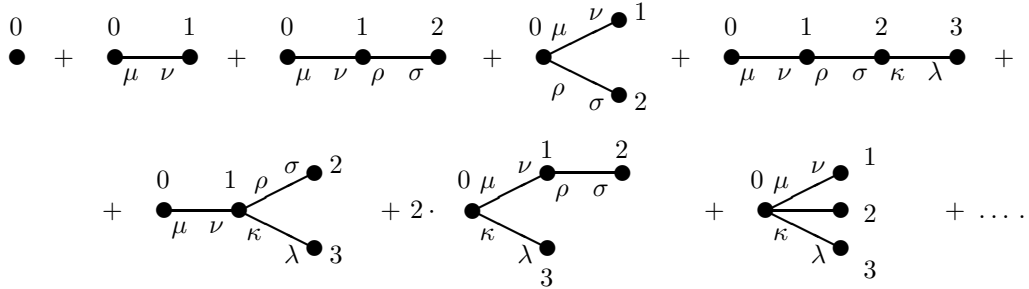
\begin{center}
\unitlength1cm
\begin{picture}(15,2)
   \put(0,0.1){\setlength{\unitlength}{1cm} 
     \begin{picture}(3,1)
       \put(1,0){\circle*{0.2}}
       \put(0.9,0.3){0}      
     \end{picture} }
   \put(1.6,0){$+$}
   \put(1.3,0.1){\setlength{\unitlength}{1cm}
     \begin{picture}(4,1)
       \put(1,0){\circle*{0.2}}
       \put(2,0){\circle*{0.2}}
       {\thicklines\put(1,0){\line(1,0){1}}}
       \put(0.9,0.3){0}\put(1.9,0.3){1}
       \put(1.1,-0.3){$\mu$}\put(1.6,-0.3){$\nu$}
     \end{picture} }
   \put(3.9,0){$+$}
   \put(3.6,0.1){\setlength{\unitlength}{1cm} 
     \begin{picture}(5,1)
       \put(1,0){\circle*{0.2}}
       \put(2,0){\circle*{0.2}}
       \put(3,0){\circle*{0.2}}
       {\thicklines\put(1,0){\line(1,0){2}}}
       \put(0.9,0.3){0}\put(1.9,0.3){1}\put(2.9,0.3){2}
       \put(1.1,-0.3){$\mu$}\put(1.6,-0.3){$\nu$}
       \put(2.1,-0.3){$\rho$}\put(2.6,-0.3){$\sigma$}
     \end{picture} }
   \put(7.3,0){$+$}
   \put(7,0.1){\setlength{\unitlength}{1cm} 
     \begin{picture}(4,2)
       \put(1,0){\circle*{0.2}}
       \put(2,-0.5){\circle*{0.2}}
       \put(2,0.5){\circle*{0.2}}
       {\thicklines\put(1,0){\line(2,1){1}}
	 \put(1,0){\line(2,-1){1}}}
       \put(0.8,0.3){0}\put(2.2,0.5){1}\put(2.2,-0.7){2}
       \put(1.1,0.3){$\mu$}\put(1.6,0.5){$\nu$}
       \put(1.1,-0.5){$\rho$}\put(1.6,-0.7){$\sigma$}
     \end{picture} }
   \put(9.8,0){$+$}
   \put(9.5,0.1){\setlength{\unitlength}{1cm}
     \begin{picture}(6,1)
       \put(1,0){\circle*{0.2}}
       \put(2,0){\circle*{0.2}}
       \put(3,0){\circle*{0.2}}
       \put(4,0){\circle*{0.2}}
       {\thicklines\put(1,0){\line(1,0){3}}}
       \put(0.9,0.3){0}\put(1.9,0.3){1}
       \put(2.9,0.3){2}\put(3.9,0.3){3}
       \put(1.1,-0.3){$\mu$}\put(1.6,-0.3){$\nu$}
       \put(2.1,-0.3){$\rho$}\put(2.6,-0.3){$\sigma$}
       \put(3.1,-0.3){$\kappa$}\put(3.6,-0.3){$\lambda$}
       \end{picture} }
   \put(14.1,0){$+$}
\end{picture}
\vspace{1.5cm}
\begin{picture}(13.5,2)
   \put(1.5,0){$+$}
   \put(1.2,0.1){\setlength{\unitlength}{1cm}
     \begin{picture}(5,1)
       \put(1,0){\circle*{0.2}}
       \put(2,0){\circle*{0.2}}
       \put(3,-0.5){\circle*{0.2}}
       \put(3,0.5){\circle*{0.2}}
       {\thicklines\put(1,0){\line(1,0){1}}
       \put(2,0){\line(2,1){1}}
       \put(2,0){\line(2,-1){1}}}
       \put(0.9,0.3){0}\put(1.8,0.3){1}
       \put(3.2,0.5){2}\put(3.2,-0.7){3}
       \put(1.1,-0.3){$\mu$}\put(1.6,-0.3){$\nu$}
       \put(2.2,0.3){$\rho$}\put(2.6,0.5){$\sigma$}
       \put(2.1,-0.4){$\kappa$}\put(2.6,-0.7){$\lambda$}
     \end{picture}  }
   \put(5.2,0){$+\,\,2\,\cdot$}
   \put(5.3,0.1){\setlength{\unitlength}{1cm}
     \begin{picture}(5,2)
       \put(1,0){\circle*{0.2}}
       \put(2,-0.5){\circle*{0.2}}
       \put(2,0.5){\circle*{0.2}}
       \put(3,0.5){\circle*{0.2}}
       {\thicklines\put(1,0){\line(2,-1){1}}
         \put(1,0){\line(2,1){1}}
	 \put(2,0.5){\line(1,0){1}}  }
       \put(0.8,0.3){0}\put(1.9,0.7){1}
       \put(2.9,0.7){2}\put(1.9,-1){3}
       \put(1.1,0.3){$\mu$}\put(1.6,0.5){$\nu$}
       \put(2.1,0.2){$\rho$}\put(2.6,0.2){$\sigma$}
       \put(1.1,-0.4){$\kappa$}\put(1.6,-0.7){$\lambda$}
     \end{picture}  }
   \put(9.5,0){$+$}
   \put(9.2,0.1){\setlength{\unitlength}{1cm}
     \begin{picture}(5,2)
       \put(1,0){\circle*{0.2}}
       \put(2,-0.5){\circle*{0.2}}
       \put(2,0){\circle*{0.2}}
       \put(2,0.5){\circle*{0.2}}
       {\thicklines
	 \put(1,0){\line(1,0){1}}
	 \put(1,0){\line(2,1){1}}
	 \put(1,0){\line(2,-1){1}}  }
       \put(0.8,0.3){0}\put(2.3,0.6){1}
       \put(2.3,-0.15){2}\put(2.3,-0.9){3}
       \put(1.1,0.3){$\mu$}\put(1.6,0.5){$\nu$}
       %\put(1.1,0.1){$\rho$}\put(1.6,0.1){$\sigma$}
       \put(1.1,-0.4){$\kappa$}\put(1.6,-0.7){$\lambda$}
     \end{picture}  }
   \put(12.7,0){$+\,\,\dots\,.$} 
\end{picture} 
\end{center}
\vspace{-5mm}
A bullet with subscript `$j$' and $l$ legs $\alpha_1,\dots,\alpha_l$
stands for the expression $(-E^{\alpha_1\cdots\alpha_l}_j)/l!$,
whereas a solid line with the indices `$\mu$' and `$\nu$' joining two bullets
`$i$' and `$j$' represents $G_{ij}^{\mu\nu}$, i.e., the Green's function
$G^{\mu\nu}({\bf x}_i-{\bf x}_j)$, plus an integration over ${\bf
x}_j\in\Omega$. The factor 2 in front of the penultimate term is the symmetry
factor which accounts for the fact that the expansion gives the two equivalent
diagrams\\ 
\vspace{-1.5cm}
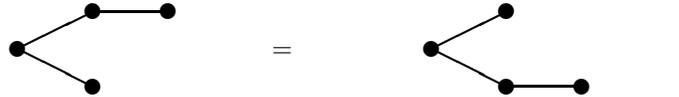
\begin{center}
\unitlength1cm
\begin{picture}(10,2)
  \put(0,0.1){\setlength{\unitlength}{1cm}
     \begin{picture}(5,2)
       \put(1,0){\circle*{0.2}}
       \put(2,-0.5){\circle*{0.2}}
       \put(2,0.5){\circle*{0.2}}
       \put(3,0.5){\circle*{0.2}}
       {\thicklines
	 \put(1,0){\line(2,-1){1}}
         \put(1,0){\line(2,1){1}}
	 \put(2,0.5){\line(1,0){1}}  }
     \end{picture}  }
  \put(4.5,0){$=$}
  \put(5.5,0.1){\setlength{\unitlength}{1cm}
     \begin{picture}(5,2)
       \put(1,0){\circle*{0.2}}
       \put(2,-0.5){\circle*{0.2}}
       \put(2,0.5){\circle*{0.2}}
       \put(3,-0.5){\circle*{0.2}}
       {\thicklines
	 \put(1,0){\line(2,-1){1}}
         \put(1,0){\line(2,1){1}}
	 \put(2,-0.5){\line(1,0){1}}  }
     \end{picture}  }
  \put(9.9,0){.}
\end{picture}
\end{center}
\par\vspace{1cm}
In a similar way, we expand the second term on the rhs.\ of
Eq.~(\ref{RGEpin}), making use of Eq.~(\ref{mincond}):
$$-\frac{1}{2}w^\alpha ({\bf x}_0)\,C\nabla^2 w^\alpha ({\bf
x}_0)=-\frac{1}{2}w^\alpha ({\bf x}_0)\,E_{(1)}^\alpha ({\bf x}_0,{\bf
u}_{(2)}({\bf x}_0)+{\bf w}({\bf x}_0))=\dots\,.$$ 
Adding the series for $E_{(1)}({\bf x}_0,{\bf u}_{(1)}({\bf x}_0))$ and
$-\frac{1}{2}w^\alpha ({\bf x}_0)\,C\nabla^2 w^\alpha ({\bf x}_0)$ we finally obtain
\vspace{-1cm}
\begin{center}
\begin{minipage}[b]{15cm}
\unitlength1cm
\begin{picture}(16,2)
   \put(0,0){$-\bar{E}_0=$}
   \put(0.5,0.1){\setlength{\unitlength}{1cm} 
     \begin{picture}(3,1)
       \put(1,0){\circle*{0.2}}
     \end{picture} }
   \put(1.9,0){$+\;\;\frac{1}{2}$}
   \put(1.9,0.1){\setlength{\unitlength}{1cm} 
     \begin{picture}(4,1)
       \put(1,0){\circle*{0.2}}
       \put(2,0){\circle*{0.2}}
       {\thicklines\put(1,0){\line(1,0){1}}}
     \end{picture} }
   \put(4.3,0){$+\;\;\frac{1}{2}$}
   \put(4.3,0.1){\setlength{\unitlength}{1cm} 
     \begin{picture}(5,1)
       \put(1,0){\circle*{0.2}}
       \put(2,0){\circle*{0.2}}
       \put(3,0){\circle*{0.2}}
       {\thicklines\put(1,0){\line(1,0){2}}}
     \end{picture} }
   \put(7.8,0){$+\;\;\frac{1}{2}$}
   \put(7.8,0.1){\setlength{\unitlength}{1cm} 
     \begin{picture}(6,1)
       \put(1,0){\circle*{0.2}}
       \put(2,0){\circle*{0.2}}
       \put(3,0){\circle*{0.2}}
       \put(4,0){\circle*{0.2}}
       {\thicklines\put(1,0){\line(1,0){3}}}
     \end{picture} }
   \put(12.4,0){$+\;\;\frac{1}{2}$}
   \put(12.4,0.1){\setlength{\unitlength}{1cm}
     \begin{picture}(5,1)
       \put(1,0){\circle*{0.2}}
       \put(2,0){\circle*{0.2}}
       \put(3,-0.5){\circle*{0.2}}
       \put(3,0.5){\circle*{0.2}}
       {\thicklines\put(1,0){\line(1,0){1}}
       \put(2,0){\line(2,1){1}}
       \put(2,0){\line(2,-1){1}}}
       \end{picture} }
\end{picture}
\vspace{1cm}
\begin{picture}(15.5,2)
   \put(1.4,0){$-\;\;\frac{1}{2}$}
   \put(1.4,0.1){\setlength{\unitlength}{1cm}
     \begin{picture}(3,2)
       \put(1,0){\circle*{0.2}}\put(2,-0.5){\circle*{0.2}}
       \put(2,0){\circle*{0.2}}\put(2,0.5){\circle*{0.2}}
       {\thicklines\put(1,0){\line(1,0){1}}
       \put(1,0){\line(2,1){1}}\put(1,0){\line(2,-1){1}}  }
     \end{picture}  }
   \put(4.4,0){$+\;\;\frac{1}{2}$}
   \put(4.4,0.1){\setlength{\unitlength}{1cm} 
     \begin{picture}(6,1)
       \put(1,0){\circle*{0.2}}\put(2,0){\circle*{0.2}}
       \put(3,0){\circle*{0.2}}\put(4,0){\circle*{0.2}}
       \put(5,0){\circle*{0.2}}
       {\thicklines\put(1,0){\line(1,0){4}}}
     \end{picture} }
   \put(10.2,0){$+\;\;\frac{1}{2}$}
   \put(10.2,0.1){\setlength{\unitlength}{1cm} 
     \begin{picture}(5,2)
       \put(1,0){\circle*{0.2}}\put(2,0){\circle*{0.2}}
       \put(3,0){\circle*{0.2}}\put(4,-0.5){\circle*{0.2}}
       \put(4,0.5){\circle*{0.2}}
       {\thicklines\put(1,0){\line(1,0){2}}
       \put(3,0){\line(2,1){1}}\put(3,0){\line(2,-1){1}}} 
     \end{picture} }
\end{picture} 
\vspace{0cm}
\begin{picture}(14.5,1.2)
   \put(1.4,0){$+$}
   \put(0.9,0.1){\setlength{\unitlength}{1cm}
     \begin{picture}(5,2)
       \put(1,0){\circle*{0.2}}\put(2,0){\circle*{0.2}}\put(3,-0.5){\circle*{0.2}}
       \put(3,0.5){\circle*{0.2}}\put(4,0.5){\circle*{0.2}}
       {\thicklines\put(1,0){\line(1,0){1}}\put(2,0){\line(2,1){1}}
       \put(2,0){\line(2,-1){1}}\put(3,0.5){\line(1,0){1}}  }
       \end{picture}  }
   \put(5.3,0){$+\;\frac{1}{2}$}
   \put(5.1,0.1){\setlength{\unitlength}{1cm}
     \begin{picture}(5,2)
       \put(1,0){\circle*{0.2}}\put(2,0){\circle*{0.2}}\put(3,-0.5){\circle*{0.2}}
       \put(3,0.5){\circle*{0.2}}\put(3,0){\circle*{0.2}}
       {\thicklines\put(1,0){\line(1,0){1}}\put(2,0){\line(2,1){1}}
       \put(2,0){\line(2,-1){1}}\put(2,0){\line(1,0){1}}  }
       \end{picture}  }
   \put(8.8,0){$-\;\frac{3}{2}$}
   \put(8.7,0.1){\setlength{\unitlength}{1cm} 
     \begin{picture}(4,2)
       \put(1,0){\circle*{0.2}}\put(2,-0.5){\circle*{0.2}}\put(2,0){\circle*{0.2}}
       \put(2,0.5){\circle*{0.2}}\put(3,0.5){\circle*{0.2}}
       {\thicklines\put(1,0){\line(1,0){1}}\put(1,0){\line(2,1){1}}
       \put(1,0){\line(2,-1){1}}\put(2,0.5){\line(1,0){1}}  }
     \end{picture} }
   \put(12.1,0){$-$}
   \put(11.7,0.1){\setlength{\unitlength}{1cm} 
     \begin{picture}(5,2)
       \put(1,0){\circle*{0.2}}\put(2,0.7){\circle*{0.2}}
       \put(2,0.3){\circle*{0.2}}\put(2,-0.3){\circle*{0.2}}
       \put(2,-0.7){\circle*{0.2}}
       {\thicklines\put(1,0){\line(3,2){1}}
       \put(1,0){\line(3,1){1}}\put(1,0){\line(3,-1){1}}
       \put(1,0){\line(3,-2){1}} }
     \end{picture} }
   \put(14,0){$+\dots\,.$}
\end{picture} 
\end{minipage}
\parbox{1cm}{\begin{equation}\label{E2_1L}\end{equation}}
\end{center}
\vspace{1cm}

Let us now turn to the energy-energy correlation function,
$\langle\bar{E}_0\bar{E}'_{0'}\rangle$, where ${\bf x}_{j'}:={\bf x}'_j$.
We write both energy functions in an expansion as done above. Using the
linearity of the averaging procedure, we obtain a series of $n$-point
correlators of the form $\langle E^{(')\alpha_1\beta_1\cdots}\cdots
E^{(')\alpha_n\beta_n\cdots}\rangle$. Assuming Gaussian disorder, we can make
use of Wick's theorem: $n$-point functions with odd $n$ vanish and those with
even $n$ can be decomposed into a sum of products of 2-point correlators. For
instance, application of Wick's theorem transforms the second-order term 
$$G_{01}^{\mu\nu}\,G_{0'1'}^{\rho\sigma}\,\langle
E_0^{\mu} E_1^{\nu} E_{0'}^{'\rho} E_{1'}^{'\sigma}\rangle$$
into the expression 
$$G_{01}^{\mu\nu}\,G_{0'1'}^{\rho\sigma}\,\Big[\langle
E_0^{\mu} E_{0'}^{'\rho}\rangle\langle E_1^{\nu} E_{1'}^{'\sigma}\rangle
+\langle E_0^{\mu} E_{1'}^{'\sigma}\rangle\langle E_1^{\nu}
E_{0'}^{'\rho}\rangle 
+\langle E_0^{\mu} E_1^{\nu}\rangle\langle E_{0'}^{'\rho}
E_{1'}^{'\sigma}\rangle\Big]\,,$$ 
diagrammatically represented by
\vspace{-0.5cm}
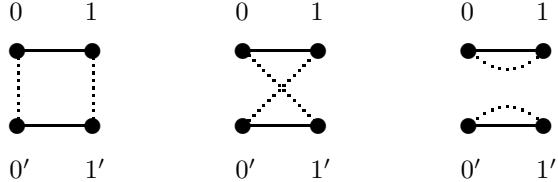
\begin{center}
\unitlength1cm
\begin{picture}(8,2.5)
\put(0,0){\setlength{\unitlength}{1cm}
  \begin{picture}(2,2)
    \put(0,1){\circle*{0.2}}\put(1,1){\circle*{0.2}} 
    \put(0,0){\circle*{0.2}}\put(1,0){\circle*{0.2}}
    {\thicklines\put(0,1){\line(1,0){1}}\put(0,0){\line(1,0){1}} }
    {\thicklines\bezier{10}(0,1)(0,0.5)(0,0)
    \bezier{10}(1,1)(1,0.5)(1,0) }
    \put(-0.1,1.4){0}\put(0.9,1.4){1} 
    \put(-0.1,-0.7){$0'$}\put(0.9,-0.7){$1'$}
  \end{picture} }
\put(3,0){\setlength{\unitlength}{1cm}
  \begin{picture}(2,2)
    \put(0,1){\circle*{0.2}}\put(1,1){\circle*{0.2}} 
    \put(0,0){\circle*{0.2}}\put(1,0){\circle*{0.2}}
    {\thicklines\put(0,1){\line(1,0){1}}\put(0,0){\line(1,0){1}} }\
    {\thicklines\bezier{15}(0,1)(0.5,0.5)(1,0)
    \bezier{15}(0,0)(0.5,0.5)(1,1) }
    \put(-0.1,1.4){0}\put(0.9,1.4){1} 
    \put(-0.1,-0.7){$0'$}\put(0.9,-0.7){$1'$}
  \end{picture} }
\put(6,0){\setlength{\unitlength}{1cm}
  \begin{picture}(2,2)
    \put(0,1){\circle*{0.2}}\put(1,1){\circle*{0.2}} 
    \put(0,0){\circle*{0.2}}\put(1,0){\circle*{0.2}}
    {\thicklines\put(0,1){\line(1,0){1}}\put(0,0){\line(1,0){1}} }
    {\thicklines\bezier{10}(0,1)(0.5,0.5)(1,1)
    \bezier{10}(0,0)(0.5,0.5)(1,0) }
    \put(-0.1,1.4){0}\put(0.9,1.4){1} 
    \put(-0.1,-0.7){$0'$}\put(0.9,-0.7){$1'$}
  \end{picture} }
\put(7.9,0.3){$.$}
\end{picture}
\end{center}
\vspace{0.5cm}
Dotted lines stand for 2-point correlation functions; the number of these
lines indicates the order in $\delta_p$ of the corresponding diagram.

The number of diagrams is considerably reduced by the special properties of
the correlations. First, the correlator $K_{R_1}({\bf u})$, being an even
function in ${\bf u}$, has all its odd derivatives vanishing at ${\bf u}=0$,
$K_{R_1}^{\alpha_1\cdots\alpha_l}(0)=0\; (l=\mbox{odd})$. Therefore, each term
containing an odd correlator connecting two points `$i$' and `$j$' on the same
`tree' (i.e., already linked by a {\em solid} line) vanishes,
\begin{eqnarray*}\langle E_i^{\alpha_1\cdots\alpha_l}\,E_j^{\beta_1\cdots\beta_n}\rangle
 &=& (-1)^n\,\delta^4({\bf x}_i-{\bf x}_j)\,K_{R_1}^{\alpha_1\cdots\alpha_l\beta_1\cdots\beta_n}({\bf
u}_{(2)}({\bf x}_i)-{\bf u}_{(2)}({\bf x}_j))\\
&=& (-1)^n\,K_{R_1}^{\alpha_1\cdots\alpha_l\beta_1\cdots\beta_n}(0)=0\quad (l+n=\mbox{odd})\,.\end{eqnarray*}
Furthermore, terms with a correlation between two {\em neighboring} points
`$i$' and `$j$' joined by a solid line, $\langle
E_i^{\alpha\cdots}\,E_j^{\beta\cdots}\rangle\propto\delta^4 ({\bf x}_i-{\bf x}_j)$,
give no contribution, because the delta function has no overlap with the
domain of integration $\Omega$. An example is given by the last diagram
above. The remaining non-vanishing diagrams up to second order in $\delta_p$
(zero and one-loop diagrams) are 
\vspace{-0.7cm}
\begin{center}
\begin{minipage}[b]{12cm}
\unitlength1cm
\begin{picture}(10,2)
\put(0,0){$\langle \bar{E}_0\,\bar{E}'_{0'}\rangle=$}
\put(2.3,-0.4){\setlength{\unitlength}{1cm}\begin{picture}(2,1.5)
  \put(0,0){\circle*{0.2}}\put(0,1){\circle*{0.2}}
  {\thicklines\bezier{10}(0,0)(0,0.5)(0,1)  }
  \end{picture} }
\put(2.6,0){$+\;\;\frac{1}{4}$}
\put(3.7,-0.4){\setlength{\unitlength}{1cm}\begin{picture}(2,1.5)
  \put(0,1){\circle*{0.2}}\put(1,1){\circle*{0.2}} 
  \put(0,0){\circle*{0.2}}\put(1,0){\circle*{0.2}}
  {\thicklines\put(0,1){\line(1,0){1}}\put(0,0){\line(1,0){1}}  }
  {\thicklines\bezier{10}(0,0)(0,0.5)(0,1)\bezier{10}(1,0)(1,0.5)(1,1) }
  \end{picture} }
\put(5.2,0){$+\;\;\frac{1}{4}$}
\put(6.4,-0.4){\setlength{\unitlength}{1cm}\begin{picture}(2,1.5)
  \put(0,1){\circle*{0.2}}\put(1,1){\circle*{0.2}} 
  \put(0,0){\circle*{0.2}}\put(1,0){\circle*{0.2}}
  {\thicklines\put(0,1){\line(1,0){1}}\put(0,0){\line(1,0){1}}  }
  {\thicklines\bezier{15}(0,0)(0.5,0.5)(1,1)\bezier{15}(0,1)(0.5,0.5)(1,0) }
  \end{picture} }
\put(8,0){$+\;\;2\cdot\frac{1}{2}$}
\put(9.5,-0.4){\setlength{\unitlength}{1cm}\begin{picture}(3,1.5)
  \put(0,1){\circle*{0.2}}\put(1,1){\circle*{0.2}} 
  \put(2,1){\circle*{0.2}}\put(0,0){\circle*{0.2}}
  {\thicklines\put(0,1){\line(1,0){2}} }
  {\thicklines\bezier{25}(0,1)(1,1.8)(2,1)\bezier{15}(0,0)(0.5,0.5)(1,1) }
  \end{picture} }
\put(11.6,0){$.$}
\end{picture}
\end{minipage}
\parbox{1cm}{\begin{equation}\label{diag1L}\end{equation}}
\end{center}
\vspace{0.7cm} 
The factor 2 in front of the last term counts the number of equivalent
contributions that appear when multiplying the two power series for the
energies. Note that so far no special assumption about $G^{\mu\nu}$ has been
made. If the matrix of Green's functions is symmetric,
$G^{\mu\nu}=G^{\nu\mu}$, the second and third diagrams in Eq.~(\ref{diag1L})
are equal.  

The energy correlator $K_{R_2}({\bf u}({\bf x}_0)-{\bf u}'({\bf
x}_0))=:\bar{K}_{00}$ at scale $R_2$ is obtained by
integrating Eq.~(\ref{diag1L}) over ${\bf x}'_0$:
$$\bar{K}_{00}=\int d^4 x'_0\,\langle\bar{E}_0\,\bar{E}'_{0'}\rangle\,.$$
By way of illustration we discuss the calculation for the second diagram in
Eq.~(\ref{diag1L}), 
$$\int d^4 x'_0\,[D2]=-K^{\mu\rho}_{00}\,\int_\Omega
d^4 x_1\,G^{\mu\nu}({\bf x}_0-{\bf x}_1)\,G^{\rho\sigma}({\bf x}_0-{\bf
x}_1)\,K^{\nu\sigma}_{11}\,.$$ 
With $R_1<|{\bf x}_0-{\bf x}_1|<R_2$ and ${\bf u}_{(2)}$ smooth over $R_1$, we
can replace $K^{\nu\sigma}_{11}$ by $K^{\nu\sigma}_{00}$ and extract it from
under the integral. 
Inserting the explicit expression for the Green's function,
Eq.~(\ref{Greenstat}), and going over to spherical coordinates, one easily
carries out the remaining integral:
\begin{equation}I_1:=\int_\Omega d^4x_1\,G({\bf x}_0-{\bf x}_1)^2 = 
-\frac{S_3}{(2\pi)^4\,C^2}\,\int_{R^2_1}^{R^2_2}\frac{dR^2}{2R^2} =
I\,\ln\left(\frac{R^2_2}{R^2_1}\right)\,,\label{I1}\end{equation} 
where $S_3=2\pi^2$ is the surface of the unit sphere in 4 dimensions and
$I:=S_3/(2(2\pi)^4\,C^2)$. The other diagrams are calculated in a similar
way. Collecting all the terms, one finds
\begin{equation}
K_{R_2}({\bf u})=K_{R_1}({\bf
u})+I\,\left(\frac{1}{2}\,K^{\mu\rho}_{R_1}({\bf u})\,K^{\mu\rho}_{R_1}({\bf
u})-K^{\mu\rho}_{R_1}({\bf
u})\,K^{\mu\rho}_{R_1}(0)\right)\,\ln\left(\frac{R^2_2}{R^2_1}\right)\,,\label{Kbar1L}\end{equation}
which yields the functional RG equation for the pinning energy correlator,
\begin{equation}\frac{\partial K_R({\bf u})}{\partial\ln
R^2}=I\,\left(\frac{1}{2}\,K_R^{\mu\rho}({\bf u})\,K_R^{\mu\rho}({\bf
u})-K_R^{\mu\rho}({\bf
u})\,K_R^{\mu\rho}(0)\right)\,,\label{FRG1L}\end{equation}
the initial condition being given by the function $K_\xi ({\bf u})$, Eq.~(\ref{K}).
This result was first obtained by Fisher\cite{f,bf} for the (4+1)-model by
means of the replica formalism. 

Eq.~(\ref{FRG1L}) is equivalent to a system of coupled differential equations
for the expansion coefficients $K_R^{\alpha_1\cdots\alpha_{2n}}(0)$ (recall
that all odd correlators $K_R^{\alpha_1\cdots\alpha_{2n+1}}(0)=0$):
\begin{eqnarray}\frac{d K_R(0)}{d\ln R^2} &=& -\frac{1}{2}\,I\,K_R^{\alpha\beta}(0)\,K_R^{\alpha\beta}(0)\,,\nonumber\\
\frac{dK_R^{\alpha\beta}(0)}{d\ln R^2} &=& 0\,,\nonumber\\
\frac{dK_R^{\alpha\beta\gamma\delta}(0)}{d\ln R^2} &=& I\,\left(K_R^{\alpha\beta\mu\rho}(0)\,K_R^{\gamma\delta\mu\rho}(0)+K_R^{\alpha\gamma\mu\rho}(0)\,
K_R^{\beta\delta\mu\rho}(0)+K_R^{\alpha\delta\mu\rho}(0)\,K_R^{\beta\gamma\mu\rho}(0)\right)\,,\label{RGK41L}\\
 &\dots& \nonumber\end{eqnarray}
The first two coefficients are readily given by  
\begin{eqnarray*}
K_R(0) &=& K_\xi(0)\left(1-\frac{1}{2}N\delta_p\,\ln\frac{R^2}{\xi^2}\right)\,,\\
K_R^{\alpha\beta}(0) &=&
K_\xi^{\alpha\beta}(0)=K_\xi(0)\frac{1}{\xi^2}\delta^{\alpha\beta}\,,\end{eqnarray*}
where $\delta^{\alpha\beta}\delta^{\alpha\beta}=N$ and Eqs.~(\ref{dercorr}) and
(\ref{K2}) have been used. Eq.~(\ref{RGK41L}) can be simplified, if one assumes
that the tensorial structure of $K_R^{\alpha\beta\gamma\delta}(0)$ is
conserved under a RG transformation. In analogy with Eq.~(\ref{K4}), we define
$\Gamma_R$ by
$K_R^{\alpha\beta\gamma\delta}(0)=:\Gamma_R\Delta^{\alpha\beta\gamma\delta}$.
Replacing this expression into Eq.~(\ref{RGK41L}), setting $\gamma=\alpha$ and
$\delta=\beta$, and summing over all indices, one finds after some algebra
\begin{equation}\frac{d\Gamma_R}{d\ln
R^2}=(N+8)\,I\,\Gamma_R^2=:A_1\,\Gamma_R^2\,,\label{RGGamma1L}\end{equation}
with the initial condition
$\Gamma_\xi=\gamma_U\xi^{N-4}\left(\frac{\pi}{2}\right)^{\frac{N}{2}}$. This
equation is easily integrated,
\begin{equation}\frac{\Gamma_R}{\Gamma_\xi}=\frac{1}{1-A_1\,\Gamma_\xi\,\ln(R^2/\xi^2)}\,,\label{Gammasol1L}\end{equation}
the solution being valid for $\xi\le R<R_c$. $\Gamma_R$ diverges at the collective
pinning radius 
\begin{equation}
R_c=\xi\,\exp\left(\frac{1}{2A_1\,\Gamma_\xi}\right)=\xi\,\exp\left(\frac{8\pi^2}{N+8}\,\left(\frac{2}{\pi}\right)^{\frac{N}{2}}\frac{1}{\delta_p}\right)\,.\label{Rc1L}\end{equation}
At that scale $K_R({\bf u})$ stops being an analytic function of ${\bf u}$
and for larger scales the system is no longer in a unique ground state. One
may show by induction that all the higher derivatives,
$K^{\alpha_1\cdots\alpha_{2n}}_R(0),\,n=3,4,\dots,$ -- if nonzero -- exhibit
a singularity at the same scale $R_c$. The one-loop RG result for $R_c$,
Eq.~(\ref{Rc1L}), confirms the lowest-order perturbation calculation,
Eq.~(\ref{Rclop}), except for the argument of the exponential, which is
smaller (for $N\le 8$) by a factor $2N/(N+8)$. 
Note that because the renormalization is arbitrarily started at the length
scale $\xi$, the present RG method cannot fix the constant of proportionality
in Eq.~(\ref{Rc1L}) just as the lowest-order perturbation calculation in
section III.A. However, with this approach there is no unknown constant as in
the condition $\langle u^2(R_c)\rangle\simeq\xi^2$; the numerical factor
inside the exponential function depends only on $\Gamma_\xi$, i.e.,
the precise choice for the form factor, Eq.~(\ref{formfactor}). 
Another advantage of this definition is that it provides a possibility for
computing higher-order corrections to $R_c$ as will be done in the next paragraph.

\subsection{Two-loop FRG equation}

The number of diagrams inflates with increasing order of $\delta_p$. There were
only 3 non-vanishing second-order (one-loop) terms, there are already 20
non-vanishing third-order (two-loop) diagrams. With the assumption of
symmetric Green's functions, $G^{\mu\nu}=G^{\nu\mu}$, only 14 topologically
different diagrams are left. Two different integral expressions occur; they
are of the form  
\begin{eqnarray*}I_2 &=& \int_{\Omega_1}dx_1\,G({\bf x}_0-{\bf
x}_1)^2\,\int_{\Omega_2}dx_2\,G({\bf x}_1-{\bf x}_2)^2\,,\\
I_3 &=& \int_{\Omega'_1}dx_1\,G({\bf x}_0-{\bf
x}_1)^2\,\int_{\Omega'_2}dx_2\,G({\bf x}_0-{\bf x}_2)\,G({\bf x}_1-{\bf
x}_2)\,,\end{eqnarray*} 
where the integration domains are such that $|{\bf x}_0-{\bf x}_1|$, $|{\bf
x}_0-{\bf x}_2|$, and $|{\bf x}_1-{\bf x}_2|$ are between $R_1$ and $R_2$.  
Accordingly, the two-loop diagrams can be regrouped into 2 classes; they are
shown in Figs.~1 and 2. The factor in front of the diagrams enumerates the
occurrence of this term in the series expansion.
\vspace{0cm}
\begin{center}
\setlength{\unitlength}{1cm}
\begin{picture}(11,3)
% D4
\put(0,0){\begin{picture}(6,3)
  \put(1,2){\circle*{0.2}}\put(2,2){\circle*{0.2}}\put(3,2){\circle*{0.2}} 
  \put(4,2){\circle*{0.2}}\put(5,2){\circle*{0.2}}\put(1,1){\circle*{0.2}} 
  {\thicklines\put(1,2){\line(1,0){4}} }
  {\thicklines\bezier{35}(1,2)(3,3.6)(5,2)\bezier{20}(2,2)(3,2.8)(4,2)
    \bezier{20}(1,1)(2,1.5)(3,2) }  
  \end{picture}  } 
% D5
\put(6,0){\begin{picture}(5,3)
  \put(1,2){\circle*{0.2}}\put(2,2){\circle*{0.2}}\put(3,2){\circle*{0.2}}
  \put(4,2.5){\circle*{0.2}}\put(4,1.5){\circle*{0.2}}\put(1,1){\circle*{0.2}} 
  {\thicklines\put(1,2){\line(1,0){2}}\put(3,2){\line(2,1){1}}
    \put(3,2){\line(2,-1){1}} }
  {\thicklines\bezier{20}(1,2)(2,2.8)(3,2)\bezier{10}(4,2.5)(4,2)(4,1.5)
    \bezier{15}(1,1)(1.5,1.5)(2,2) }
  \end{picture}  }
\end{picture}
\begin{picture}(14,3)
% D6
\put(0,0){\setlength{\unitlength}{1cm}\begin{picture}(5,3)
  \put(1,2){\circle*{0.2}}\put(2,2){\circle*{0.2}}\put(3,1.5){\circle*{0.2}} 
  \put(3,2.5){\circle*{0.2}}\put(4,2.5){\circle*{0.2}}\put(1,1){\circle*{0.2}} 
  {\thicklines\put(1,2){\line(1,0){1}}\put(2,2){\line(2,1){1}} 
    \put(2,2){\line(2,-1){1}}\put(3,2.5){\line(1,0){1}} }
  {\thicklines\bezier{20}(1,2)(2,1.2)(3,1.5)\bezier{20}(2,2)(3,3.6)(4,2.5)
    \bezier{20}(1,1)(2,1.3)(3,2.5) }
  \put(0.2,1.4){$2$} \end{picture} }
% D7
\put(5,0){\setlength{\unitlength}{1cm}\begin{picture}(4,3)
  \put(1,2){\circle*{0.2}}\put(2,2){\circle*{0.2}}\put(3,2.5){\circle*{0.2}} 
  \put(3,2){\circle*{0.2}}\put(3,1.5){\circle*{0.2}}\put(1,1){\circle*{0.2}} 
  {\thicklines\put(1,2){\line(1,0){1}}\put(2,2){\line(2,1){1}} 
    \put(2,2){\line(2,-1){1}}\put(2,2){\line(1,0){1}} }
  {\thicklines\bezier{20}(1,2)(2,3)(3,2.5)\bezier{5}(3,2)(3,1.75)(3,1.5)
    \bezier{15}(1,1)(1.5,1.5)(2,2) }
  \put(0.2,1.4){$3$} \end{picture} }
% D8
\put(10,0){\setlength{\unitlength}{1cm}\begin{picture}(4,3)
  \put(1,2){\circle*{0.2}}\put(2,2.5){\circle*{0.2}}\put(2,2){\circle*{0.2}} 
  \put(2,1.5){\circle*{0.2}}\put(3,2.5){\circle*{0.2}}\put(1,1){\circle*{0.2}} 
  {\thicklines\put(1,2){\line(2,1){1}}\put(1,2){\line(1,0){1}} 
    \put(1,2){\line(2,-1){1}}\put(2,2.5){\line(1,0){1}} }
  {\thicklines\bezier{20}(1,2)(1.7,3.3)(3,2.5)\bezier{5}(2,2)(2,1.75)(2,1.5)
    \bezier{25}(1,1)(3.4,1)(2,2.5) }
  \put(0,1.4){$-3$} \end{picture} }
\end{picture}
\begin{picture}(14,3)
% D9
\put(0,0){\begin{picture}(3,4)
  \put(1,2){\circle*{0.2}}\put(2,3){\circle*{0.2}}\put(2,2.5){\circle*{0.2}}
  \put(2,1.5){\circle*{0.2}} \put(2,1){\circle*{0.2}}\put(1,1){\circle*{0.2}} 
  {\thicklines\put(1,2){\line(1,1){1}}\put(1,2){\line(2,1){1}}
    \put(1,2){\line(2,-1){1}}\put(1,2){\line(1,-1){1}} }
  {\thicklines\bezier{5}(2,3)(2,2.75)(2,2.5)\bezier{5}(2,1.5)(2,1.25)(2,1)
    \bezier{10}(1,2)(1,1.5)(1,1) }
  \put(0,1.4){$-6$} \end{picture} }
% D10
\put(5,0){\setlength{\unitlength}{1cm}\begin{picture}(4,3)
  \put(1,2){\circle*{0.2}}\put(2,2){\circle*{0.2}}\put(3,1.5){\circle*{0.2}} 
  \put(3,2.5){\circle*{0.2}}\put(1,1){\circle*{0.2}}\put(2,1){\circle*{0.2}}
  {\thicklines\put(1,2){\line(1,0){1}}\put(2,2){\line(2,1){1}}
    \put(2,2){\line(2,-1){1}}\put(1,1){\line(1,0){1}} }
  {\thicklines\bezier{10}(1,1)(1,1.5)(1,2)\bezier{10}(2,1)(2,1.5)(2,2)
    \bezier{10}(3,1.5)(3,2)(3,2.5) }
  \end{picture} }
% D11
\put(10,0){\setlength{\unitlength}{1cm}\begin{picture}(4,3)
  \put(1,2){\circle*{0.2}}\put(2,2){\circle*{0.2}}\put(3,1.5){\circle*{0.2}} 
  \put(3,2.5){\circle*{0.2}}\put(1,1){\circle*{0.2}}\put(2,1){\circle*{0.2}}
  {\thicklines\put(1,2){\line(1,0){1}}\put(2,2){\line(2,1){1}}
    \put(2,2){\line(2,-1){1}}\put(1,1){\line(1,0){1}} }
  {\thicklines\bezier{20}(1,2)(2,3)(3,2.5)\bezier{15}(1,1)(1.5,1.5)(2,2)
    \bezier{10}(2,1)(2.5,1.25)(3,1.5) }
  \put(0.2,1.4){$2$} \end{picture} }
\end{picture}
\begin{picture}(14,3)
% D12
\put(0,0){\setlength{\unitlength}{1cm}\begin{picture}(4,3)
  \put(1,2){\circle*{0.2}}\put(2,2){\circle*{0.2}}\put(3,2){\circle*{0.2}}
  \put(1,1){\circle*{0.2}}\put(2,1){\circle*{0.2}}\put(3,1){\circle*{0.2}}
  {\thicklines\put(1,2){\line(1,0){2}}\put(1,1){\line(1,0){2}} }
  {\thicklines\bezier{10}(1,1)(1,1.5)(1,2)\bezier{10}(2,1)(2,1.5)(2,2)
    \bezier{10}(3,1)(3,1.5)(3,2) }
  \put(0.2,1.4){$\frac{1}{2}$} \end{picture} }
% D13
\put(5,0){\setlength{\unitlength}{1cm}\begin{picture}(4,3)
  \put(1,2){\circle*{0.2}}\put(2,2){\circle*{0.2}}\put(3,2){\circle*{0.2}}
  \put(1,1){\circle*{0.2}}\put(2,1){\circle*{0.2}}\put(3,1){\circle*{0.2}}
  {\thicklines\put(1,2){\line(1,0){2}}\put(1,1){\line(1,0){2}} }
  {\thicklines\bezier{20}(1,2)(2,2.8)(3,2)\bezier{10}(2,1)(2,1.5)(2,2)
    \bezier{20}(1,1)(2,0.2)(3,1) }
  \put(0.2,1.4){$\frac{1}{4}$} \end{picture} }
% D14
\put(10,0){\setlength{\unitlength}{1cm}\begin{picture}(4,3)
  \put(1,2){\circle*{0.2}}\put(2,2.5){\circle*{0.2}}\put(2,2){\circle*{0.2}} 
  \put(2,1.5){\circle*{0.2}}\put(1,1){\circle*{0.2}}\put(2,1){\circle*{0.2}} 
  {\thicklines\put(1,2){\line(2,1){1}}\put(1,2){\line(1,0){1}} 
    \put(1,2){\line(2,-1){1}}\put(1,1){\line(1,0){1}} }
  {\thicklines\bezier{5}(2,2.5)(2,2.25)(2,2)\bezier{5}(2,1.5)(2,1.25)(2,1)
    \bezier{10}(1,2)(1,1.5)(1,1) }
  \put(0,1.4){$-3$} \end{picture} }
\end{picture}
\end{center}
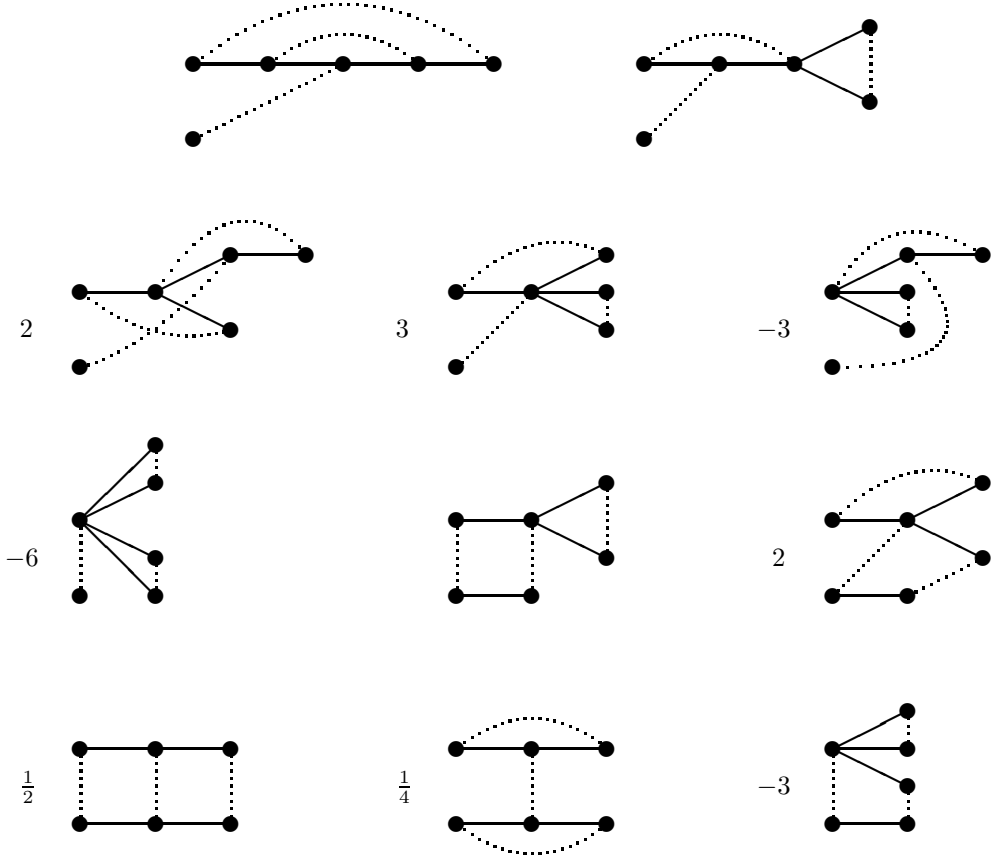
\begin{center}\parbox{8cm}{{\small Fig.~1: Two-loop diagrams in the class
$I_2$.}}\end{center} 
\vspace{0.5cm}
\begin{center}
\setlength{\unitlength}{1cm}
\begin{picture}(14,3)
% D15
\put(0,0){\setlength{\unitlength}{1cm}\begin{picture}(4,3)
  \put(1,2){\circle*{0.2}}\put(2,2){\circle*{0.2}}\put(3,2){\circle*{0.2}}
  \put(1,1){\circle*{0.2}}\put(2,1){\circle*{0.2}}\put(3,1){\circle*{0.2}}
  {\thicklines\put(1,2){\line(1,0){2}} \put(1,1){\line(1,0){2}} }
  {\thicklines\bezier{10}(1,1)(1,1.5)(1,2)\bezier{10}(2,1)(2.5,1.5)(3,2)
    \bezier{10}(2,2)(2.5,1.5)(3,1) }
  \put(0.2,1.4){$\frac{1}{2}$} \end{picture} }
% D16
\put(5,0){\setlength{\unitlength}{1cm}\begin{picture}(4,3)
  \put(1,2){\circle*{0.2}}\put(2,2){\circle*{0.2}}\put(3,2){\circle*{0.2}}
  \put(1,1){\circle*{0.2}}\put(2,1){\circle*{0.2}}\put(3,1){\circle*{0.2}}
  {\thicklines\put(1,2){\line(1,0){2}} \put(1,1){\line(1,0){2}} }
  {\thicklines\bezier{10}(1,1)(1.5,1.5)(2,2)
    \bezier{10}(2,1)(2.5,1.5)(3,2)\bezier{20}(1,2)(2,1.5)(3,1) }
  \put(0.2,1.4){$\frac{1}{2}$} \end{picture} }
% D17
\put(10,0){\setlength{\unitlength}{1cm}\begin{picture}(5,3)
  \put(1,2){\circle*{0.2}}\put(2,2){\circle*{0.2}}\put(3,2){\circle*{0.2}}
  \put(4,2){\circle*{0.2}}\put(1,1){\circle*{0.2}}\put(2,1){\circle*{0.2}}
  {\thicklines\put(1,2){\line(1,0){3}}\put(1,1){\line(1,0){1}} }
  {\thicklines\bezier{25}(1,2)(2.5,3)(4,2)\bezier{10}(1,1)(1.5,1.5)(2,2)
    \bezier{10}(2,1)(2.5,1.5)(3,2) }
  \end{picture} }
\end{picture}
\end{center}
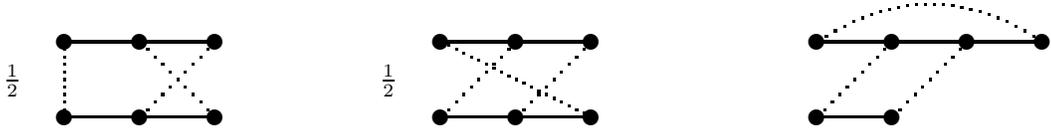
\begin{center}\parbox{8cm}{{\small Fig.~2: Two-loop diagrams in the class
$I_3$.}}\end{center} 
Recalling that a $l$-leg bullet carries a factor $1/l!$, we calculate these
diagrams as demonstrated in the previous paragraph. The third-order
contributions to $K_{R_2}$, Eq.~(\ref{Kbar1L}), are found to be
\begin{eqnarray} 
K_{R_2}^{2-loop}({\bf u}) &=& \left[\frac{1}{2}\left(K_{R_1}^{\mu\rho}({\bf u})-K_{R_1}^{\mu\rho}(0)\right)\,\left(K_{R_1}^{\kappa\tau}({\bf u})-K_{R_1}^{\kappa\tau}(0)\right)\,K_{R_1}^{\mu\rho\kappa\tau}({\bf u})\right]\,I_2\nonumber\\
&+&
\Big[\left(K_{R_1}^{\mu\rho}({\bf u})-K_{R_1}^{\mu\rho}(0)\right)\,K_{R_1}^{\kappa\tau\mu}({\bf u})\,K_{R_1}^{\kappa\tau\rho}({\bf u})\Big]\,I_3\,.\label{Kbar2L}\end{eqnarray}
Here, we have already taken into account the diagonal structure of the Green's
matrix, $G^{\mu\nu}=\delta^{\mu\nu}\,G$. In section III.B we have seen that
the relevant quantity for determining $R_c$ is $\Gamma_R$, which is
essentially the fourth derivative of $K({\bf u})$. We therefore differentiate
Eqs.~(\ref{Kbar1L}) and (\ref{Kbar2L}) twice with respect to $u^\alpha$ and
twice with respect to $u^\beta$, set ${\bf u}=0$ and use the definition of
$\Gamma_R$,
$K_R^{\alpha\alpha\beta\beta}(0)=\Gamma_R\Delta^{\alpha\alpha\beta\beta}=\Gamma_R
N(N+2)$. After summing over all indices, we eventually obtain
\begin{equation}\Gamma_{R_2}=\Gamma_{R_1}+\Gamma_{R_1}(N+8)I_1+\Gamma_{R_1}^2
(N^2+6N+20)I_2+\Gamma_{R_1}^3 (20N+88)I_3\,.\label{Gammabar2L}\end{equation}

Let us compute the two integrals $I_2$ and $I_3$. 
$I_2$ is found at once, cf.~Eq.~(\ref{I1}), 
\begin{equation}I_2=(I_1)^2=I^2\,\ln^2\left(\frac{R_2^2}{R_1^2}\right)\,.\label{I2}\end{equation}
The calculation of $I_3$ is less trivial. It can be written in the form
$$I_3=\frac{1}{(2\pi)^8\,C^2}\,\int_{\Omega_1}\frac{d^4x}{x^4}\,\int_{\Omega_2}\frac{d^4y}{y^2\,
({\bf x}+{\bf y})^2}\,,$$
where the integration domains $\Omega_1$ and $\Omega_2$ are such that $|{\bf x}|,
|{\bf y}|$, and $|{\bf x}+{\bf y}|$ are taken between $R_1$ and $R_2$. This
integral is well known from $\phi^4$-field theory\cite{b}. In the limit
$R_1\longrightarrow 0\,,\;R_2\longrightarrow\infty$, it is logarithmically
divergent, the divergent contributions being of the form $\ln(R_2/R_1)$ and
$\ln^2(R_2/R_1)$. To extract these terms, we take $R_2\gg R_1$, so that the
condition $R_1<|{\bf x}+{\bf y}|<R_2$ can be omitted with good
approximation. We introduce spherical coordinates as usual, $R:=|{\bf x}|,
r:=|{\bf y}|$, and choose the $y_4$-axis parallel to ${\bf x}$:
$$I_3\approx\frac{4\pi\,S_3}{(2\pi)^8\,C^2}\,\int_{R_1^2}^{R_2^2}\frac{dR^2}{2R^2}\,\int_{R_1^2}^{R_2^2}
\frac{dr^2}{2}\,\int_0^\pi\frac{\sin^2\vartheta\,d\vartheta}{R^2+r^2+2Rr\cos\vartheta}\,,\quad
R_2\gg R_1\,.$$
The $dr^2$-integration is split into two parts, $R_1^2<r^2<R^2$ and
$R^2<r^2<R_2^2$; for each part the $d\vartheta$-integration is carried out
with help of the method of residua by transforming to coordinates
$z:=\exp(i\vartheta)$. The remaining integrals pose no problem and one obtains
$$I_3=I^2\,\left[\ln\left(\frac{R_2^2}{R_1^2}\right)+\frac{1}{2}\ln^2\left(\frac{R_2^2}{R_1^2}\right)+
const.+O\left(\frac{R_1^2}{R_2^2}\right)\right]\,.$$
The change $R_1\longrightarrow R_2$ need not be infinitesimal. We are free to
choose $R_2\gg R_1$ in such a way that $\ln(R_2/R_1)\gg 1$, as long as
$\delta_p\,\ln(R_2/R_1)\ll 1$, so that $\Gamma_{R_2}-\Gamma_{R_1}$ remains
small. For weak pinning, $\delta_p\ll 1$, this condition is fulfilled and the
last two terms in the above expression can be neglected,
\begin{equation}I_3\approx I^2\,\left[\ln\left(\frac{R_2^2}{R_1^2}\right)+\frac{1}{2}\ln^2\left(\frac{R_2^2}{R_1^2}
\right)\right]\,.\label{I3}\end{equation}

Inserting the expressions for $I_2$ and $I_3$, Eqs.~(\ref{I2}) and
(\ref{I3}) into Eq.~(\ref{Gammabar2L}), we obtain the logarithmic corrections up
to two-loop approximation: 
\begin{equation}\frac{\Gamma_{R_2}}{\Gamma_{R_1}}=1+A_1\Gamma_{R_1}\ln\left(\frac{R_2^2}{R_1^2}\right)+A_1^2\Gamma_{R_1}^2\ln^2\left(\frac{R_2^2}{R_1^2}\right)+A_2\Gamma_{R_1}\ln\left(\frac{R_2^2}{R_1^2}\right)\,,\label{Gammasum2L}\end{equation}
with $A_1=(N+8) I$ and $A_2=(20N+88) I^2$. The fact that the {\em two}-loop term
$\propto\ln^2$ is the exact square of the {\em one}-loop contribution
$(A_1\Gamma_{R_1}\ln)$ provides a check for the correctness of the one-loop RG
equation, Eq.~(\ref{RGGamma1L}). Indeed, from Eq.~(\ref{Gammasol1L}) one
deduces
$$\frac{\Gamma_{R_2}}{\Gamma_{R_1}}=1+A_1\Gamma_{R_1}\ln\left(\frac{R_2^2}{R_1^2}\right)+A_1^2\Gamma_{R_1}^2\ln^2\left(\frac{R_2^2}{R_1^2}\right)+O(\delta_p^3)\,;$$
this shows that the $\ln^2$-contribution in Eq.~(\ref{Gammasum2L}) is already
taken into account in the one-loop equation and need therefore not be
considered explicitly. The two-loop RG equation for $\Gamma_R$ then reads:
\begin{equation}\frac{d\Gamma_R}{d\ln
R^2}=A_1\,\Gamma_R^2+A_2\,\Gamma_R^3\,.\label{RGGamma2L}\end{equation} 
Similarly, the two-loop FRG equation for $K_R({\bf u})$ is inferred from
Eqs.~(\ref{Kbar1L}) and (\ref{Kbar2L}), by omitting all the $\ln^2$-terms in
$I_2$ and $I_3$:
\begin{eqnarray}\frac{\partial K_R({\bf u})}{\partial\ln
R^2} &=& I\left(\frac{1}{2}\,K_R^{\mu\rho}({\bf u})\,K_R^{\mu\rho}({\bf
u})-K_R^{\mu\rho}({\bf u})\,K_R^{\mu\rho}(0)\right)\\ \nonumber
 & &+\frac{I^2}{2}\Bigg( K_R^{\mu\rho}({\bf
u})-K_R^{\mu\rho}(0)\Bigg) K_R^{\kappa\tau\mu}({\bf
u})\,K_R^{\kappa\tau\rho}({\bf u})\,.\label{FRG2L}\end{eqnarray}

The integration of the differential equation (\ref{RGGamma2L}),
\begin{eqnarray*}\int_{\ln\xi^2}^{\ln R^2}d\ln R'^2 & = &
\int_{\Gamma_\xi}^{\Gamma_R}\frac{d\Gamma}{A_1\,\Gamma^2+A_2\,\Gamma^3}\,,\\
& = &
\int_{\Gamma_\xi}^{\Gamma_R}\,d\Gamma\,\left[\frac{1}{A_1\,\Gamma^2}-\frac{A_2}{A_1^2\,\Gamma}+\frac{A_2}{A_1^2\,(\Gamma+A_1/A_2)}\right]\,,\end{eqnarray*}
yields the solution
$$\ln\left(\frac{R^2}{\xi^2}\right)=-\frac{1}{A_1\,\Gamma_R}+\frac{1}{A_1\,\Gamma_\xi}+\frac{A_2}{A_1^2}\ln\left(1+\frac{A_1}{A_2\,\Gamma_R}\right)-\frac{A_2}{A_1^2}\ln\left(1+\frac{A_1}{A_2\,\Gamma_\xi}\right)\,.$$
To extract $R_c$, we take the limit $\Gamma_R\rightarrow\infty$ and use the
assumption of weak pinning, so that $A_2\Gamma_\xi/A_1\propto\delta_p\ll 1$:
$$\ln\left(\frac{R_c^2}{\xi^2}\right)=\frac{1}{A_1\,\Gamma_\xi}+\frac{A_2}{A_1^2}\,\ln\left(\frac{A_2\,\Gamma_\xi}{A_1}\right)\,.$$
Solving with respect to $R_c$ and inserting the explicit expressions for $A_1$,
$A_2$, $I$, and $\Gamma_\xi$, leads to the final result for the collective
pinning radius,
\begin{equation}
R_c\simeq\xi\,\delta_p^{\alpha_2}\,\exp\left(\frac{\alpha_1}{\delta_p}\right)\,,\label{Rc2L}\end{equation}
where
$$\alpha_1=\frac{8\pi^2}{N+8}\left(\frac{2}{\pi}\right)^{\frac{N}{2}}\quad\mbox{and}\quad\alpha_2=\frac{2(5N+22)}{(N+8)^2}\,.$$
Comparison with the one-loop result, Eq.~(\ref{Rc1L}), shows that the two-loop
calculation produces an algebraic correction with exponent $\alpha_2$ (the
constant $\alpha_1$ in the exponential function remains unchanged).
Finally, scaling arguments provide us with an improved result for the critical
force density, 
$$F_c\simeq\frac{C}{\xi}\left(\frac{\xi}{R_c}\right)^2=\frac{C}{\xi}\,\delta_p^{-2\alpha_2}\,\exp\left(-\frac{2\alpha_1}{\delta_p}\right)\,.$$

\subsection{Dynamic approach and RG} 

In this paragraph we apply the above results to the dynamic approach
introduced in Sec.~III.B. We investigate the behavior of the
friction coefficient $\eta(v)=\eta_\circ+\delta\eta(v)$ of Eq.~(\ref{flow})
under the renormalization group and arrive at an estimate for the critical force
density $F_c$ in an alternative way.

In order to pave the way for our real-space RG, we rewrite Eq.~(\ref{dyn1})
through the Fourier representation of the Green's function, cf.\
Eq.~(\ref{Greendyn}), 
$$v^\alpha\,\delta\eta=\int_0^\infty dt\,K^{\alpha\beta\beta}_\xi({\bf
v}t)\,\int\frac{d^4k}{(2\pi)^4}\,\frac{1}{\eta}\,\exp\left(-\frac{C k^2
t}{\eta}\right)\,,$$ 
where we have consistently replaced $\eta_\circ$ by $\eta$.
Since disorder is weak, the critical velocity $v_c$ is expected to be
small. Hence, we expand the correlator $K^{\alpha\beta\beta}_\xi({\bf v}t)$ into a 
power series around ${\bf v}t=0$ and retain the first non-vanishing term,
proportional to $\Gamma_\xi$: 
$$\delta\eta\simeq 
(N+2)\,\frac{\Gamma_\xi}{\eta}\,\int\frac{d^4k}{(2\pi)^4}\int_0^\infty
dt\,t\,\exp\left(-\frac{C k^2 t}{\eta}\right)\,.$$ 
The time and angular integrations are easily carried out,
$$\delta\eta=(N+2)\,\Gamma_\xi\,\eta\,I\int\frac{dk^2}{k^2}\,.$$
This form is appropriate to make contact to our results. Let us derive a RG
equation for the viscosity: Assume we have renormalized $\eta$ up to a scale
$R_1>\xi$. Going over to the larger scale $R_2$, the viscosity is increased by
\begin{eqnarray*}\eta_{R_2}-\eta_{R_1} &=& (N+2)\,I\,\Gamma_{R_1}\,\eta_{R_1}\int_{1/R_2^2}^{1/R_1^2}\frac{dk^2}{k^2}\,,\\
 &=&
(N+2)\,I\,\Gamma_{R_1}\,\eta_{R_1}\ln\left(\frac{R_2^2}{R_1^2}\right)\,.\end{eqnarray*}
As a result, the desired one-loop RG equation for the viscosity reads
\begin{equation}\frac{d\eta_R}{d\ln
R^2}=(N+2)\,I\,\Gamma_R\,\eta_R\,.\label{RGeta1L}\end{equation}
The corresponding formula for the (4+1)-model has recently been derived by
Leschhorn {\em et al.}\cite{nstl,lnst}, using essentially the same
method. Eq.~(\ref{RGeta1L}) is solved by separation of variables,
\begin{equation}\int_{\eta_\circ}^{\eta_R}\frac{d\eta}{\eta}=(N+2)\,I\,\int_{\Gamma_\xi}^{\Gamma_R}d\Gamma\,\Gamma\,\frac{d\ln
R^2}{d\Gamma}\,.\label{etasol}\end{equation}
Use of the one-loop renormalization flow of $\Gamma_R$,
Eqs.~(\ref{RGGamma1L}) and (\ref{Gammasol1L}), yields
\begin{equation}1+\frac{\delta\eta_R}{\eta_\circ}=\frac{\eta_R}{\eta_\circ}=\left(\frac{\Gamma_R}{\Gamma_\xi}\right)^\sigma=\left(A_1\Gamma_\xi\ln\frac{R_c^2}{R^2}\right)^{-\sigma}\,,\label{etasol1L}\end{equation}
with $\sigma=(N+2)/(N+8)$. 

In section III.B we defined the critical force $F_c$ as the applied force
$F_{ext}$ for which $\delta\eta(F_{ext})\simeq\eta_\circ$, reflecting the onset
of nonlinearity in the velocity-force characteristics as $F_{ext}$ is
lowered. Here, we employ an alternative criterion, which we believe is better
adapted to the problem. If a (finite) critical force $F_c>0$ exists, then
$\lim_{F_{ext}\rightarrow F_c^+}v(F_{ext})=0$. As a consequence, 
$$\lim_{F_{ext}\rightarrow F_c^+}\eta(F_{ext})=\lim_{F_{ext}\rightarrow
F_c^+}\frac{F_{ext}}{v(F_{ext})}\longrightarrow\infty$$
and we can define $F_c$ via the singularity of the effective
viscosity $\eta=\eta_\circ+\delta\eta$. Obviously, $\eta_R$ diverges for
$R\rightarrow R_c$, the critical force $F_c$ is thus related to the
collective pinning radius $R_c$. From scaling we know that
\begin{equation}F_{ext}\simeq\frac{C}{\xi}\frac{\xi^2}{R^2}\,.\label{Fscal}\end{equation}
Recalling Eq.~(\ref{Rc1L}), we thus find
\begin{equation}F_c\simeq\frac{C}{\xi}\frac{\xi^2}{R_c^2}=\frac{C}{\xi}\,\exp\left(-\frac{16\pi^2}{N+8}\,\left(\frac{2}{\pi}\right)^{\frac{N}{2}}\frac{1}{\delta_p}\right)\,,\label{Fc1L}\end{equation}
which essentially agrees with the lowest-order perturbation result
Eq.~(\ref{Fcdynap}), the argument of the exponential function being smaller by
a factor $\sigma=(N+2)/(N+8)$. As usual, the number in front of the exponential
remains unspecified. With the help of the scaling relation (\ref{Fscal}), the
viscosity, see Eq.~(\ref{etasol1L}), can be written as a function of $F_{ext}$,
\begin{equation}\frac{\eta(F_{ext})}{\eta_\circ}=\left(A_1\Gamma_\xi\,\ln\frac{F_{ext}}{F_c}\right)^{-\sigma}=\left[1-\frac{N+8}{16\pi^2}\,\left(\frac{\pi}{2}\right)^{\frac{N}{2}}\delta_p\,\ln\left(\frac{C}{\xi
F_{ext}}\right)\right]^{-\sigma}\,,\label{eta1L}\end{equation} 
providing us with an explicit form of the velocity-force characteristics
$v=F_{ext}/\eta(F_{ext})$ versus $F_{ext}$.
Whereas the lowest-order perturbation result, Eq.~(\ref{etalop}), is only
valid for applied forces $F_c\ll F_{ext}<C/\xi$, this expression holds down to
much smaller forces and thus better describes the non-linear regime. Expanding
Eq.~(\ref{eta1L}) for small $\delta_p$ and comparing the resulting series with
Eq.~(\ref{etalop}), one can check that both expressions coincide (for
$F_{ext}\ll C/\xi$) up to first order in $\delta_p$ as they ought to.

\section{Summary and Conclusion}

Using a RG approach, we have calculated the collective pinning radius $R_c$ as
well as the critical force density $F_c$ for a (4+$N$)-dimensional elastic
manifold subject to point defects. We have confined ourselves to the case of
zero temperature and have assumed weak and short-range disorder of
Gaussian type.

By means of a real-space renormalization procedure a functional RG equation
for the pinning energy correlator $K_R({\bf u})$ was found in one and
two-loop approximation. It has been shown that the fourth derivative of that
correlator taken at zero ${\bf u}$,
$K_R^{\alpha\beta\gamma\delta}(0)=\Gamma_R\Delta^{\alpha\beta\gamma\delta}$,
possesses a singularity at some finite scale of renormalization which we have
identified with the collective pinning radius $R_c$. To one-loop correction we
have derived the expression $R_c=\alpha_0\xi\exp(\alpha_1/\delta_p)$,
$\alpha_1=(2/\pi)^{N/2}\,8\pi^2/(N+8)$, Eq.~(\ref{Rc1L}), that confirms the
result obtained by a simple lowest-order perturbation calculation,
Eq.~(\ref{Rclop}), the argument of the exponential function differing by a factor
$2N/(N+8)$. In two-loop approximation we have found an additional
algebraic factor $\propto\delta_p^{\alpha_2}$ with an exponent
$\alpha_2=2(5N+22)/(N+8)^2$, see Eq.~(\ref{Rc2L}). Contrary to the
perturbation calculation, the RG approach succeeds in fixing both $\alpha_1$
and $\alpha_2$. The constant of proportionality $\alpha_0$, however, remains
undetermined in either of the two methods.     

Next, we have derived a RG equation for the viscosity $\eta$ that appears when
driving the system by an external force, cf.\ Eq.~(\ref{RGeta1L}). The
previously obtained result for $\Gamma_R$ has allowed us to solve this
equation, see Eq.~(\ref{eta1L}). In comparison with the lowest-order
perturbation result, Eq.~(\ref{etalop}), the RG solution provides us with an
improved description of the velocity-force characteristics in the non-linear
regime where the disorder potential considerably influences the vortex motion.
We have defined the critical force $F_c$ via the singularity of the viscosity
$\eta$ and indicated how $F_c$ and $R_c$ are both related. As was the case for
$R_c$, the one-loop expression we have found for $F_c$ agrees with the result
obtained by means of dimensional estimates and the dynamical approach,
Eqs.~(\ref{Fcdimest}) and (\ref{Fcdynap}) respectively, except that the number
in the exponential function is now fixed.

The improved result for $R_c$ provides a firm starting point for the scaling
analysis in the random manifold regime (large scales $R>R_c$).
In a forthcoming publication\cite{wglb}, we apply the same method to the physical
case of a three-dimensional system of vortices in type-II superconductors with
dispersive elastic moduli and gain quantitative refinements of the weak
collective pinning theory.

\section*{Acknowledgments}

The authors are grateful to G.T.~Zimanyi, L.~Balents, M.~M\'{e}zard,
T.~Giamarchi, and P.~le Doussal for valuable discussion on the subject and
gratefully acknowledge the support by the Swiss National Foundation.

%%%%%%%%%%%%%%%%%%%%%%%%%%%%%%%%%%%%%%%%%%%%%%%%%%%%%%%%%%%%%%%%%%%%%%%%%%%%%%

\end{document}